%% file: GenEl_IJoES_2023-anonym.tex
\documentclass[preprint,12pt]{elsarticle}

\usepackage{amssymb}
\usepackage{amsmath}
\usepackage{mathrsfs}
\usepackage{bbm}
\usepackage{xcolor}

\newcommand{\noi}{\noindent}
\newcommand{\be}{\begin{equation}}
\newcommand{\ee}{\end{equation}}

\journal{International Journal of Engineering Science}

\begin{document}

\begin{frontmatter}



\title{General theory for plane extensible elastica with arbitrary undeformed shape}


\author[inst1,inst2]{Alessandro Taloni}

\affiliation[inst1]{organization={CNR -- Consiglio Nazionale delle Ricerche, Istituto dei Sistemi Complessi},
            addressline={via dei Taurini 19}, 
            city={Roma},
            postcode={00185}, 
            country={Italy}}
\affiliation[inst2]{organization={Center for Complexity and Biosystems, Department of Physics, University of Milan},
            addressline={Via Giovanni Celoria, 16}, 
            city={Milan},
            postcode={20133}, 
            country={Italy}}

\author[inst3,inst4]{Daniele Vilone}

\affiliation[inst3]{organization={Laboratory of Agent Based Social Simulation, Institute of Cognitive Science and Technology, CNR},
            addressline={via Palestro, 32}, 
            city={Roma},
            postcode={00185}, 
            country={Italy}}
\affiliation[inst4]{organization={Grupo Interdisciplinar de Sistemas Complejos, Departamento de Matemáticas, Universidad Carlos III de Madrid},
            addressline={Leganés}, 
            city={Madrid},
            postcode={28911}, 
            country={Spain}}
            
\author[inst5]{Giuseppe Ruta}
\affiliation[inst5]{organization={Dipartimento di Ingegneria strutturale e geotecnica, Sapienza Università di Roma},
            addressline={Via Eudossiana, 18}, 
            city={Roma},
            postcode={00184}, 
            country={Italy}}

\begin{abstract}
A general expression for the strain energy of a homogeneous, isotropic, plane extensible elastica with an arbitrary undeformed configuration is derived. This energy constitutes the correct expression for one-dimensional models of polymers or vesicles, whose natural configuration is characterized by locally changing curvature. We derive the macroscopic stress-strain relations, providing an universal  criterion  for the neutral curve location.  In this respect, we demonstrate that  the natural curve existence constitutes the fundamental requirement for the conformational dynamics of any inextensbile biological filament.

\end{abstract}



\end{frontmatter}



\section*{Introduction} \label{sec:intro}
    

The old theories of flexure were based on the assumption that the elastica strain consists of extension and contraction of longitudinal filaments, said fibers~\cite{bernoulli1691specimen,euler1771genuina,euler1960rational}.  In this context, the reduction of the problem to the one-dimensional plane strain~\cite{michell1901theory} was naturally taken as the compelling point of view.  By instance,   Lagrange  writes about the elastica  as  a \emph{fil flexible et en m\^{e}me temps extensible et contractible}~\cite{de1853mecanique}. Yet, in Kirchoff theory the three-dimensional deformation of a slender elastic rod is reduced to the bending deformation of a one-dimensional  curve~\cite{kirchhoff1859ueber}.
Moreover, the theory of the bending and twisting of thin rods and wires was for a long time developed independently of the general three-dimensional equations of elasticity, by one-dimensional methods akin  to those employed by Bernoulli and  Euler~\cite{love2013treatise}. However,  the problem of which, among the fibers composing the elastica,  was eligible to be taken as the representative curve, remained unsettled till the profound works of Parent     and Coulomb. Already in 1695, Jacob Bernoulli had argued that the bending moment had to be taken, at each cross-section, with respect to the point where it intersects the \emph{line of fulcra}, i.e. the line which does not suffer any extension nor contraction all along the deformation: the neutral fiber. If the tensions vary linearly over the rod rigid cross-section, the neutral line coincides with its central  line. Although this fact was  already known by Beeckman, Hooke, Huygens, Varignon and Mariotte \cite{euler1960rational,timoshenko1983history}, it was only in 1713 that these observations took the mathematical form of the Parent criterion~\cite{parent1713veritable}.  

The issue of the existence and ensuing location of the neutral fiber, avoiding any specific elastic hypothesis about the law of
variation of the tensions over the cross-section, remains today unjustly unattended by modern scientists. Indeed, what was considered as \emph{the} elastic problem for a hundred of years has been regarded with condescension by later historians. Mostly because it is wrongly assumed that later works on three-dimensional elasticity somehow fixed it. However, 
in modern theories of bending, such as  Saint Venant's, the existence of the neutral line is implicitly postulated \emph{ad hoc}, not demonstrated~\cite{truesdell1984works,euler1960rational}. Our purpose, in this paper, is to tackle a problem of a more limited reach: to identify a criterion for the neutral fiber existence and placement, if the linearity of the Hooke law is assumed, but the starting configuration of the elastica is not necessarily straight, as in the slender beam classical theory, but any. 

The problem of spontaneously curved bands was already known to Jacob Bernoulli as testified by a posthumous fragment published in 1744~\cite{truesdell1984works}.  In the same year also the Euler's treatise on elastic curves appeared, in which the law of an elastica endowed with a non-zero curvature was asserted~\cite{euler1960rational}. A systematic theory of rods with curved undeformed configuration is substantially due to Clebsch~\cite{clebsch1862theorie}, although outlined by Kirchoff~\cite{kirchhoff1859ueber}, to whom probably the notion of ``unstressed state'' must be acknowledged.  Only in  Timoshenko's theory of curved beam~\cite{timoshenko1955strength}, however,  it is shown that the neutral fiber does not correspond to the line of centroids. Surprisingly, in the 20th century the problem of the elastica whose undeformed shapes are circular arcs or rings, in particular, has been  taken up by several authors~\cite{kammel1966einfluss,antman1968general,atanackovic1998buckling,lagrange2016wrinkling,vakakis1999buckling, schmidt1979critical,fu1995initial, katifori2009collapse,kosel1984biegelinie,troger2012nonlinear,chaskalovic1995bifurcation,schmidt1996discussion}, seemingly  neglecting Timoshenko's work.  As a matter of fact, when the one-dimensional representation of the naturally skew elastica is provided, usually this corresponds to the line of centroids \cite{antman1968general,atanackovic1998buckling}, assumed, sometimes tacitly, to be the neutral fiber, which can be extensible ~\cite{antman1968general,atanackovic1998buckling,kammel1966einfluss,lagrange2016wrinkling,tadjbakhsh1966variational,magnusson2001behaviour,oshri2016properties}, or inextensible~\cite{vakakis1999buckling, schmidt1979critical,fu1995initial, katifori2009collapse,kosel1984biegelinie,troger2012nonlinear,chaskalovic1995bifurcation,schmidt1996discussion}.

In this paper we consider the deformation of an elastica with a spontaneous curvature which is not constant, thus generalizing the theory of curved beams. Our work is motivated by the fact that natural biological filaments cannot be considered straight in their native state, nor endowed with a uniform curvature. To the contrary, biological materials develop into a variety of complex shapes, which can sense and respond to local curvature.  This field of mechanobiology~\cite{schoen2013yin} poses fundamental questions about the function of geometry and, particularly, on the role that the local curvature plays in biological systems~\cite{schamberger2022curvature}. In particular, the curvature appears as a determining factor in important biological functional tasks executed by  many  bio-polymers inside the cell~\cite{harada1987sliding,ghosh2014dynamics, bausch2006bottom,brangwynne2008cytoplasmic,schaller2010polar}, as well as in the design of next-generation genomics technologies \cite{dorfman2018statistical}.

On top of that, none of the three major classes of cytoskeletal filaments, microtubules, actin filaments, and intermediate filaments, can be considered homogeneous nor isotropic, as each is formed from chains of discrete protein monomers \cite{alberts2017molecular}.  At the typical level of description in usual molecular mechanics models, however, they are often treated as traditional engineering elements \cite{howard2002mechanics}:  they are considered plane elasticae whose mechanical behavior is entirely dictated by the Bernoulli-Euler theory. Our model does not deviate from this conceptual framework; however, our construction can be easily generalized to realistically account for non-homogeneous filaments with varying geometrical and mechanical properties.

Our starting point is the derivation of the strain energy  associated to any three-dimensional elastica deformation, as the energy cost of a shape fluctuation is the significant starting concept of any dynamical model in polymer physics~\cite{rubinstein2003polymer,doi1988theory}. The one-dimensional representation of the elastica arises as the natural context for the strain energy formulation (Section~\ref{sec:model}). The macroscopic constitutive relations are obtained from the strain energy function (Section~\ref{sec:const_rel}), and in Section~\ref{sec:neutral} we demonstrate that the strain energy minimization  furnishes the criterion for locating the neutral curve. We  conclude the paper in Section~\ref{sec:disc}, discussing the implication of our findings in terms of polymer dynamics and statitics, providing the evidence that the developed framework yields the correct generalization of  flexible and semiflexible polymers models, and of simplified one-dimensional
models of fluctuating vesicles. We discuss the relevance that the old-fashioned concept of neutral fiber acquires in modern polymer physics.  We show, indeed, that a renovated interest around it, is motivated by the fact that its existence is provided by the  minimum energy principle at the base of any polymer conformational change. We discuss the correct and convenient way of  implementing  our model in computer simulations and, finally, the relevance of our model in  describing real biological non-homogeneous filaments.

\section{The Model}
\label{sec:model}

\ 
We consider the deformation of an homogeneous \emph{plane extensible  elastica}~\cite{antman1968general}. This is defined as an hypereleastic three-dimensional body of rectangular section, thin in two of its dimensions (height $h$ and depth $b$), but not in its length, and subject to the only restriction that its deformations take place in a plane, thus excluding twisting. According to the Cosserat's theories \cite{cosserat1909theorie,green1974theory1,green1974theory2,rubin2000cosserat,naghdi1984constrained}, any elastic deformation stems from an undeformed body configuration or undeformed state, defined as the shape corresponding to the unstressed condition~\cite{love2013treatise}. The work performed in passing from the undeformed to the deformed state is the strain or stored energy, which is the pivotal concept in our model.

\begin{figure}[ht]
\centering
\includegraphics[width=8.0cm]{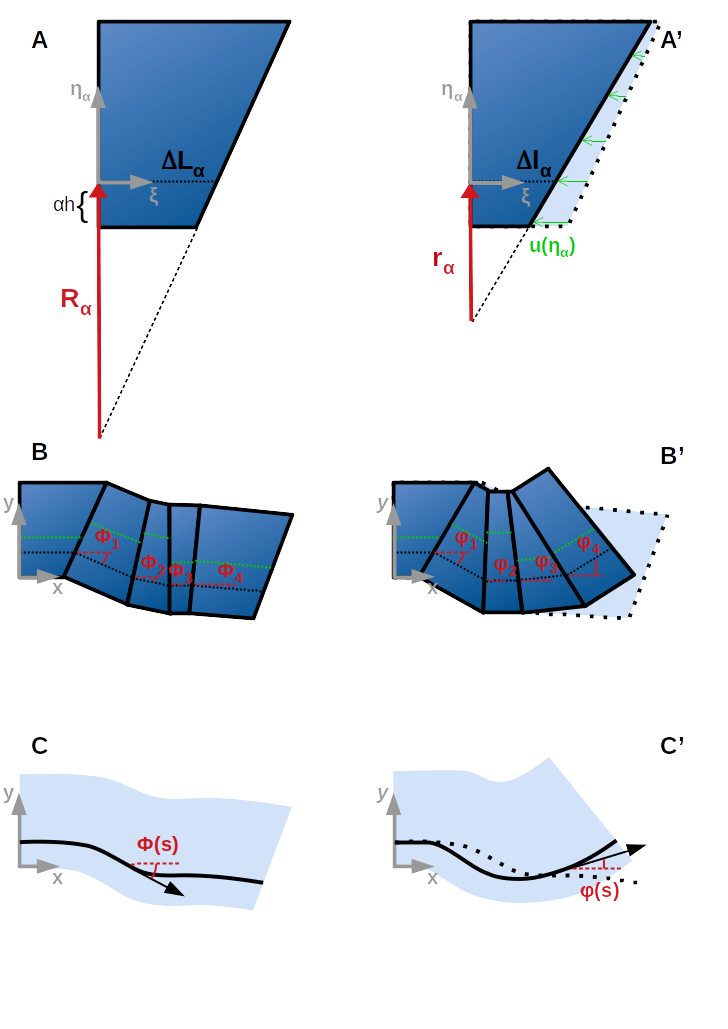}
\caption{Derivation of the strain energy.  {\bf A:} The elementary block,  with its integral system of reference $\xi\text{-}0\text{-}\eta_\alpha$, has the origin  placed at an height $\alpha h $ from the bottom surface ($\alpha\in[0,1]$). A reference fiber is characterized by its length ($\eta_\alpha=0$)  is $\Delta L_\alpha$ (dotted black line) and by $R_\alpha$, with the corresponding spontaneous  curvature $K_\alpha=\frac{1}{|R_\alpha|}$.  $R_\alpha$ is the distance of the origin of the local reference from the intersection between the extended sides of the block, with a positive sign if its orientation is concordant with that of $\eta_\alpha$.  It becomes the curvature radius in the continuum limit. {\bf A$'$:} The block undergoes a deformation to a new shape (dark blue), where the original configuration is  in light blue. Shape change entails  $R_\alpha\to r_\alpha$. Any fiber of the block endures a deformation   represented by the fiber displacement $u(\eta_\alpha)=\Delta l(\eta_\alpha) -\Delta L(\eta_\alpha) $ (green arrows). {\bf B:} The   undeformed elastica  is displayed as a sequence of elementary identical blocks, each with its own shape. The sequence of the reference segments is shown as a dotted black polygonal $L_\alpha$, while bending angle $\Phi^{(i)}$ of each block is in red. The green dashed lines correspond to the $\alpha_U^{(i)}$ fibers. Conventionally, the construction assumes that the left section of each block remains straight-angled, so that the bending angles are defined at the left side of each block. {\bf B$'$:} The deformation of the  elastica involves the deformation of the reference polygonal chain, $L_\alpha\to l_\alpha$ (dotted black line). The bending angles vary from $\Phi^{(i)}$ to $\varphi^{(i)}$. 
{\bf C:} In the continuum limit, the  material reference curve is represented by a solid black curve. The tangent to this curve, $\mathbf{T}_\alpha(s)$, forms with the $x$ axis the spontaneous bending angle $\Phi(s)$. {\bf C$'$:} The deformed material curve  (solid black line) stems from its undeformed configuration (dotted black line). The black arrow represents the new tangent $\mathbf{t}_\alpha(s)$, while its corresponding bending angle $\varphi(s)$ is in red.}
\label{fig1}
\end{figure}

We propose to compute the strain energy  using a finite difference scheme, as displayed in Fig.~\ref{fig1}. The undeformed body is schematically subdivided in $N$ elementary blocks of arbitrary volumes  $\Delta V^{(i)}$ ($i=[1,N]$) (Fig.~\ref{fig1}A). To each elementary volume is assigned a local reference system, where the $\xi$ and $\eta_\alpha$ axes define respectively the block's longitudinal and transverse directions,  and the origin is arbitrarily placed at an height $\alpha h$ ($0\leq\alpha\leq1$) from the block's bottom surface, see Fig.~\ref{fig1}A, and at the left side of the block as shown in figure. We now introduce a reference segment of length $\Delta L_\alpha^{(i)}$, equivalent to the longitudinal dimension of the plane $\eta_\alpha=0$ (Fig.~\ref{fig1}A).  Assuming the cross-sections  to remain rigid, the body deformation necessarily involves the deformation of each block, passing from the volume $\Delta V^{(i)}$ to $\Delta v^{(i)}$. At the same time, the reference segment $\Delta L_\alpha^{(i)}$ transforms into $\Delta l_\alpha^{(i)}$. As in the old theories of elastic flexure, we assume that the stress is determined by the strain~\cite{truesdell2004non}, and the strain consists of the displacements of independent longitudinal filaments  $u^{(i)}(\eta_\alpha)=\Delta l^{(i)}(\eta_\alpha)-\Delta L^{(i)}(\eta_\alpha)$. Embracing the microscopical validity of the Hooke's law, we sum, \emph{à la} Leibniz, over the contributions of the fibers upon the entire cross section. Thus,  the elementary block's strain energy can be written as

\be
\Delta E_\alpha^{(i)}=\frac{bY}{2}\int_{-\alpha h}^{(1-\alpha)h}\left[\frac{u^{(i)}(\eta_\alpha)}{\Delta L^{(i)}(\eta_\alpha)}\right]^2 \Delta L^{(i)} (\eta_\alpha)\ d\eta_\alpha , 
    \label{E_block}
\ee

\noindent where $Y$ is a parameter which has the dimensions of a stress and depends on the material properties. Therefore the strain energy~\eqref{E_block} represents the work performed in deforming the elementary block from the unstressed configuration reported in Fig.~\ref{fig1}A, to that in Fig.~\ref{fig1}A$'$. However, in continuity with the old theories,  the strain energy~\eqref{E_block} is associated to the deformation of the representative  material segment. Nevertheless, $\Delta E_\alpha^{(i)}$ is a scalar quantity and, as such, must be invariant under  change of the local reference frame, i.e.  change of material representative  segment. As a matter of fact,   by applying the transformation  $\eta_\alpha=\eta_\beta-(\alpha-\beta)h$, the property $\Delta E_\alpha^{(i)}=\Delta E_\beta^{(i)}$ is always satisfied, as it is demonstrated in the Supplementary Informations (SI). The analytical derivation of the energy~\eqref{E_block} is fully developed in SI. We hereby report the final expression:

\be
\Delta E_\alpha^{(i)} = \frac{Y}{2}\Delta L_\alpha^{(i)}\left\{F^{(i)}_\alpha\, {\varepsilon^{(i)}_\alpha}^2+2S^{(i)}_\alpha\, \varepsilon^{(i)}_\alpha\mu^{(i)}_\alpha+I^{(i)}_\alpha\, {\mu^{(i)}_\alpha}^2\right\} \ , 
    \label{E_block_any}
\ee

\noindent where we have introduced the axial strain

\be
\varepsilon_\alpha^{(i)}=\frac{\Delta l_\alpha^{(i)}-\Delta L_\alpha^{(i)}}{\Delta L_\alpha^{(i)}}
\label{axial_strain}
\ee

\noindent and the bending measure~\cite{antman1968general}

\be
\mu_\alpha^{(i)}=\frac{1}{\Delta L_\alpha^{(i)}}\left(\frac{\Delta l_\alpha^{(i)}}{r^{(i)}_\alpha}-\frac{\Delta L^{(i)}_\alpha}{R^{(i)}_\alpha}\right).
\label{bending}
\ee

\noindent The quantities $R^{(i)}_\alpha$ and $r^{(i)}_\alpha$ are the radii connecting the intersection point between the extensions  of the block limiting sections, with the origin of the block's local reference system:  they have a sign corresponding to their orientation, as shown in the SI  (see Fig.~\ref{fig1}A,A$'$ for the details). In the following we will show how, in the continuum limit, they correspond to the local radii of curvature of the undeformed and deformed configurations, respectively. For simplicity, from now on we will refer to such quantities as curvature radii also in the discrete case. Likewise, we will define the block spontaneous curvature as $K^{(i)}_\alpha=\frac{1}{\left|R^{(i)}_\alpha\right|}$, while $k^{(i)}_\alpha=\frac{1}{\left|r^{(i)}_\alpha\right|}$. 

Most importantly, according to the Timoshenko's curved beam theory~\cite{timoshenko1955strength}, $F^{(i)}_\alpha$ represents the reduced area, $I_\alpha^{(i)}$ the reduced moment of inertia and $S_\alpha^{(i)}$ is the   reduced axial-bending coupling moment. They   have a simple integral expression furnished in SI, corresponding to the formula obtained by K\"{a}mmel in the theory of a ring allowing axial compressibility and subjected to hydrostatic pressure~\cite{kammel1966einfluss}, and subsequently used in the analysis of compressible rings~\cite{atanackovic1998buckling}.  Our derivation, however,    sheds light on a fundamental aspect that seemingly remained unnoticed in the past analysis: that is, $F_\alpha^{(i)}$, $S_\alpha^{(i)}$ and $I_\alpha^{(i)}$ take different forms according to whether the  intersection between the sidelines containing the block's sections lies above or below the elementary volume, i.e. they depend on the direction of $R_\alpha^{(i)}$ (see SI). This dependence is compactly expressed as

\be
F_\alpha^{(i)} = \frac{b}{K_\alpha^{(i)}}\ln\left(1+h\,\max[K_0^{(i)},K_1^{(i)}]\right)
\label{F_alpha},
\ee

\be
S_\alpha^{(i)}=\frac{\mbox{sgn}\left(K_0^{(i)}-K_1^{(i)}\right)}{K^{(i)}_\alpha}\left[bh-F^{(i)}_\alpha\right]
\label{S_alpha},
\ee

\be
I_\alpha^{(i)}=\frac{\mbox{sgn}\left(K_0^{(i)}-K_1^{(i)}\right)}{K^{(i)}_\alpha}\left[b\left(\frac{1}{2}-\alpha\right)\,h^2-S^{(i)}_\alpha\right].
\label{I_alpha}
\ee

\noi If the center of the intersection point is placed below the block's bottom line ($\alpha=0$), hence $K_0^{(i)}>K_1^{(i)}$ and the function $\mbox{sgn}\left(K_0^{(i)}-K_1^{(i)}\right)=1$. In the opposite case, the curvature of the upper fiber $\alpha=1$ is such that $K_1^{(i)}>K_0^{(i)}$, and $\mbox{sgn}\left(K_0^{(i)}-K_1^{(i)}\right)=-1$. By inspection of the expression~\eqref{S_alpha}, there exists one value of $\alpha$ for which stretching and bending term are decoupled, i.e. $S_\alpha^{(i)}=0$. This value corresponds to

\be
\alpha_U=
\begin{cases}
\frac{1}{\ln\left(1+h\,K_0\right)}-\frac{1}{hK_0}&  K_0>K_1\\
\ \\
1-\left[\frac{1}{\ln\left(1+h\,K_1\right)}-\frac{1}{hK_1}\right]&   K_1>K_0 \ ,
\label{alpha_NF}
\end{cases}
\ee

\noindent with $S_{\alpha}>0$ for $\alpha<\alpha_U$, and $S_{\alpha}<0$ for $\alpha>\alpha_U$. Moreover, it is worth to notice that it results 
$$
\lim_{K_{0/1}\rightarrow0^+}\alpha_U = \frac{1}{2}.
$$

\noindent This means that the fiber allowing the axial-bending uncoupling coincides with the line of centroids in the case of flat undeformed configuration, recovering the classical beam theories result, as shown in the SI.

Formally, the full body strain energy is the sum of the elementary blocks strain energy, i.e. $E_\alpha=\sum_{i=1}^N \Delta E_\alpha^{(i)}$. In our finite difference scheme,  the ordered sequence of consecutive $\Delta L^{(i)}_\alpha$  constitutes the representative undeformed material polygonal chain  $L_\alpha$ in Fig.~\ref{fig1}B. Correspondingly, the ordered sequence of consecutive $\Delta l^{(i)}_\alpha$  composes the representative deformed material polygonal chain  $l_\alpha$  in  Fig.~\ref{fig1}B$'$. Now, adopting as unique reference system the laboratory frame as in Fig.\ref{fig1}B,B$'$,  the strain energy $E_\alpha$ is  expressible in terms of the  axial strain~\eqref{axial_strain}, and of the bending measure $\mu_\alpha^{(i)}$, that assumes an alternative but equivalent expression to that in~\eqref{bending} ~\cite{antman1968general,kammel1966einfluss,atanackovic1998buckling,greenberg1967equilibrium}:

\be
\mu_\alpha^{(i)}=\frac{\Delta\Phi^{(i)}-\Delta\varphi^{(i)}}{\Delta L_\alpha^{(i)}},
\label{bending_new}
\ee

\noindent where $\Phi^{(i)}$ ($\varphi^{(i)}$) is the $(i)$-th cross-sectional bending angle with respect the $x$--axis in the undeformed (deformed) configuration (see Fig.~\ref{fig1}B and ~\ref{fig1}B$'$). Hence

\be
E_\alpha = \frac{Y}{2}\sum_{i=1}^N\Delta L_\alpha^{(i)}\left\{F^{(i)}_\alpha\, {\varepsilon^{(i)}_\alpha}^2+2S^{(i)}_\alpha\, \varepsilon^{(i)}_\alpha\mu^{(i)}_\alpha+I^{(i)}_\alpha\, {\mu^{(i)}_\alpha}^2\right\} \ . 
    \label{E_block_dscrt}     
\ee

\noindent In our finite difference scheme, the blocks are assumed to be independent. Thus, nothing prevents us from assigning to each block a different local reference frame, i.e., a different $\alpha$. As a matter of fact, thanks to the property of invariance of $\Delta E_\alpha^{(i)}$ under a change of local reference frame, the total energy $E_\alpha$ is also invariant. We will revisit this important consideration in Sec.~\ref{sec:neutral} when we discuss the existence of the neutral fiber.

The continuum limit is taken by increasing $N$ up to the point that the poligonal chain $L_\alpha$ ($l_\alpha$) tends to a finite curve, namely the reference curve, or representative fiber $\mathscr{L}_\alpha$ ($\ell_\alpha$) (Fig.~\ref{fig1}C,~\ref{fig1}C$'$). According to the older theories of flexure, a material fiber can be envisaged as a translation of one of the outermost curves ($\alpha=0$ or $\alpha=1$) along the orthogonal
cross sections of the body. The Cartesian components of the material line, or fiber, in the laboratory frame are expressed as $\mathbf{L_\alpha}(s)\left(X_\alpha(s),Y_\alpha(s)\right)$ and $\mathbf{l_\alpha}(s)\left(x_\alpha(s),y_\alpha(s)\right)$, respectively, where $s$ is the same internal parameter for both  the undeformed reference fiber $\mathscr{L}_\alpha:[s_m,s_M]\to \mathbbm{R}^2$ and  deformed one $\ell_\alpha:[s_m,s_M]\to \mathbbm{R}^2$. By introducing the tangents $\mathbf{T_\alpha}=\frac{d\mathbf{L_\alpha}}{ds}$ and $\mathbf{t_\alpha}=\frac{d\mathbf{l_\alpha}}{ds}$ to $\mathscr{L}_\alpha$ and $\ell_\alpha$, the continuum limit of the  elastica strain energy is

\begin{equation*}
\mathscr{E}_\alpha = \frac{Y}{2}\int_{s_m}^{s_M}ds
\left\{\frac{F_\alpha(s)}{\left|\mathbf{T_\alpha}(s)\right|}\, \left[\left|\mathbf{t_\alpha}(s)\right|-\left|\mathbf{T_\alpha}(s)\right|\right]^2 + \right.
\end{equation*}
\be  
\left. - \frac{2S_\alpha(s)}{\left|\mathbf{T_\alpha}(s)\right|}\, \left[\left|\mathbf{t_\alpha}(s)\right|-\left|\mathbf{T_\alpha}(s)\right|\right]\left[\varphi'(s)-\Phi'(s)\right] +
\frac{I_\alpha(s)}{\left|\mathbf{T_\alpha}(s)\right|}\,\left[\varphi'(s)-\Phi'(s)\right]^2\right\},
\label{E_ct}
\ee

\noindent where $(\cdot)'$ denotes the derivative with respect to $s$.  The expression~\eqref{E_ct} costitutes one of the main results of our analysis, extending the Timoshenko's linear theory of curved beams to the case of an elastica with generic undeformed condition, identified by varying local spontaneous curvature  $K(s)=\left|\Phi'(s)\right|$.

\noindent
In SI the differential form of the quantities in~\eqref{F_alpha},~\eqref{S_alpha} and~\eqref{I_alpha} are furnished. In particular, it is worth of noticing how the condition for the axial-bending uncoupling~\eqref{alpha_NF} must be valid  locally in the continuum limit, i.e. $S_\alpha(s)=0$. This 
can be achieved only if $\alpha_U$ in~\eqref{alpha_NF} changes with $s$ according to

\be
\alpha_U(s)=
\frac{\left|\mathbf{T}_0(s)\right|}{h\Phi'(s)}-\frac{\left|\Phi'(s)\right|}{\Phi'(s)}\frac{1}{\ln\left(1+h\frac{\left|\Phi'(s)\right|}{\min[\left|\mathbf{T}_0(s)\right|,\left|\mathbf{T}_1(s)\right|]}\right)}.
\label{alpha_NF_ct}
\ee

\noindent The classical case of a  thin slender rod, beam or bar~\cite{euler1960rational}, is also derived in SI for the sake of completeness. It is shown  how this constitutes the zero curvature limit of the general model represented by the energy expression~\eqref{E_ct}.

\section{Macroscopic constitutive relations\label{sec:const_rel}}

We now derive the macroscopic stress-strain relations for the elastica, given a generic undeformed configuration.
Our starting point is the strain energy density, defined as $W_\alpha=\frac{\Delta E_\alpha}{\Delta L_\alpha}$. Therefore, when the undeformed state has a constant curvature throughout the whole elastica, from Eq.~\eqref{E_block_any} it turns out that for a generic choice of the representative fiber

\be
W_\alpha=\frac{Y}{2}\left\{F_\alpha\, {\varepsilon_\alpha}^2+2S_\alpha\, \varepsilon_\alpha\mu_\alpha+I_\alpha\, {\mu_\alpha}^2\right\}
 \label{E_function-any}.
\ee

\noindent The macroscopic constitutive equations can be drawn from~\eqref{E_function-any}, according to the derivation furnished  in~\cite{antman1968general}, which employs the definition of the elastica as a  three-dimensional
body, or to an alternative variational approach in which the elastica is treated as a one-dimensional medium~\cite{tadjbakhsh1966variational}:

\be
\begin{cases}
      N_\alpha=\frac{\partial W_\alpha}{\partial \varepsilon_\alpha}=Y\left(F_\alpha\varepsilon_\alpha+S_\alpha\mu_\alpha\right)\\
      \ \\
      M_\alpha=\frac{\partial W_\alpha}{\partial \mu_\alpha}=Y\left(S_\alpha\varepsilon_\alpha+I_\alpha\mu_\alpha\right).
    \end{cases}
    \label{const_eq-any}
\ee              

\noi $N_\alpha$ and $M_\alpha$ are the elastica axial force and bending moment respectively, where the one-dimensional representation of the elastica coincides with a generic fiber $\alpha$. We recall that the bending measure enjoys two analogue definitions,  Eq.s~\eqref{bending} and~\eqref{bending_new}. From Eq.s\eqref{const_eq-any}, it is immediately verified that linear decoupled relations only hold for the choice of the material fiber allowing for the axial-bending uncoupling of the strain energy~\eqref{E_block_any}, namely $\alpha=\alpha_U$ in Eq.\eqref{alpha_NF}. The same condition is valid in the classical case of flat  initial configuration, as it is shown in SI.

 It is instructive to see how the constitutive relations transform under change of reference line, i.e. when shifting from the fiber $\alpha$ to $\beta$. To this end,  we firstly need to express how the strain and the bending measure are modified if the material line changes:
 
\be
\varepsilon_\alpha=\frac{\Delta L_\beta}{\Delta L_\alpha}\left[\varepsilon _\beta+h(\alpha-\beta)\mu_\beta\right],
\label{ch_strain_any}
\ee

\noi and

\be
\mu_\alpha=\frac{\Delta L_\beta}{\Delta L_\alpha}\mu_\beta. 
\label{ch_bend_any}
\ee

\noindent Therefore, plugging these relations into~\eqref{const_eq-any}, it turns out that 

\be
\begin{cases}
N_\alpha=N_\beta\\
      \ \\
M_\alpha=M_\beta+h(\beta-\alpha)N_\beta.      
\end{cases}
    \label{const_eq-any_transf}
\ee    

\

\noindent The former transformations appear to be universal, in the sense that they are valid in the case of slender bars as well, as shown in SI.
Importantly, while the axial force is invariant under change of representative fiber, the bending moment gains an axial force contribution which accounts for the different curvatures of the fibers $\alpha$ and $\beta$. Moerover, when $N_\alpha=0$,  the bending moment is invariant under changes of representative fiber: $M_\alpha=M_\beta$. Physically this corresponds to the fact that, for a pure bending transformation, the bending moment cannot depend on the fiber chosen as representative.  For any other transformation for which $N_\alpha\neq 0$, the bending moment will depend on the fiber on which it is computed, and the axial force contribution has to be summed up to the total moment of the forces.

\ 

 When the curvature is locally changing within the elastica, the continuum version of the strain energy density~\eqref{E_function-any} is 

\begin{equation*}
\mathscr{W}_\alpha(s) = \frac{Y}{2\left|\mathbf{T_\alpha}(s)\right|^2}
\left\{F_\alpha(s)\, \left[\left|\mathbf{t_\alpha}(s)\right|-\left|\mathbf{T_\alpha}(s)\right|\right]^2 + \right.
\end{equation*}
\be
\left. -2S_\alpha(s)\, \left[\left|\mathbf{t_\alpha}(s)\right|-\left|\mathbf{T_\alpha}(s)\right|\right]\left[\varphi'(s)-\Phi'(s)\right] +I_\alpha(s)
\,\left[\varphi'(s)-\Phi'(s)\right]^2\right\},
\label{strain_E_func_ct_any}
\ee

\noindent from which the constitutive equations in a local form can be defined as

\be
\begin{cases}
      N_\alpha(s)=\frac{\partial\mathscr{W}_\alpha(s)}{\partial \left|\mathbf{t_\alpha}(s)\right| }=\frac{Y}{\left|\mathbf{T_\alpha}(s)\right|^2}\left\{F_\alpha(s)\left[\left|\mathbf{t_\alpha}(s)\right|-\left|\mathbf{T_\alpha}(s)\right|\right]-S_\alpha(s)\left[\varphi'(s)-\Phi'(s)\right]\right\}\\
      \ \\
      M_\alpha(s)=\frac{\partial\mathscr{W}_\alpha(s)}{\partial \varphi'(s)}=\frac{Y}{\left|\mathbf{T_\alpha}(s)\right|^2}\left\{-S_\alpha(s)\left[\left|\mathbf{t_\alpha}(s)\right|-\left|\mathbf{T_\alpha}(s)\right|\right]+I_\alpha(s)\left[\varphi'(s)-\Phi'(s)\right]\right\}.
    \end{cases}
    \label{const_eq_ct_any}
\ee

\noi Once again, the value of $\alpha$ needs to be varied throughout the elastica according to the law~\eqref{alpha_NF_ct}, for the stress-strain relationships~\eqref{const_eq_ct_any} to be locally decoupled.

\section{Neutral fiber location\label{sec:neutral}}

\begin{figure}[ht]
\centering
\includegraphics[width=8.1cm]{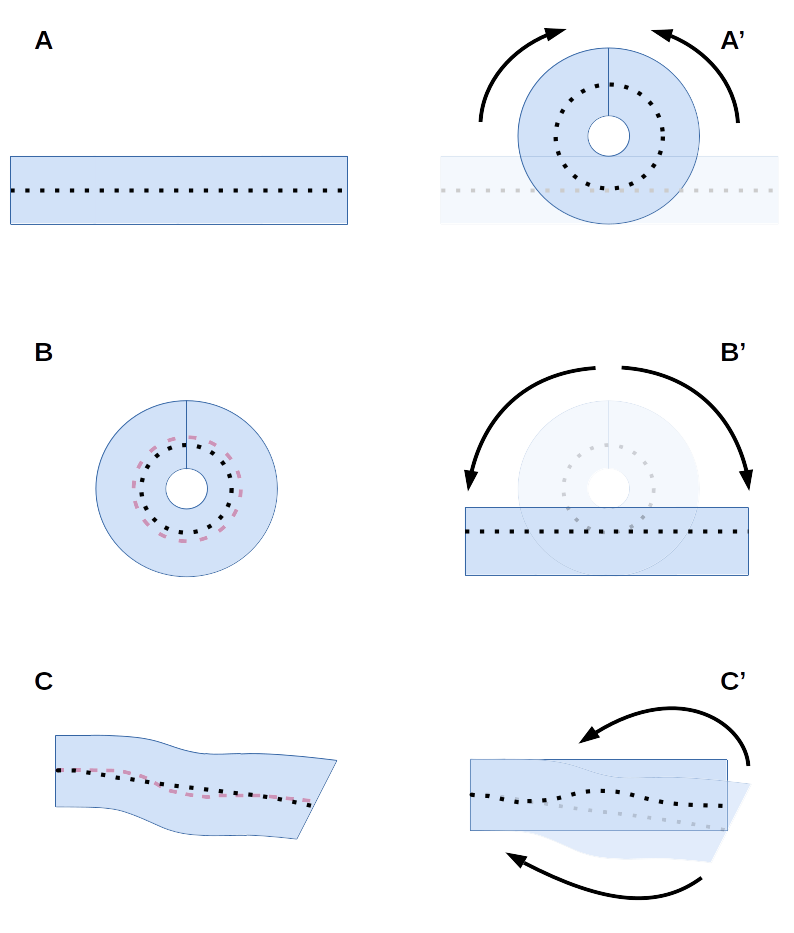}
\caption{Neutral fiber location. {\bf A:} The undeformed condition coincides with a straight bar of length $L$. The line of centroids ($\alpha=1/2$) is depicted as a dotted black fiber: taking this as the representative one-dimensional medium would give the axial-bending uncoupling (see SI). {\bf A$'$:} The beam in panel A is bent into a ring. If the line of centroids maintains  constant its length, therefore this coincides with the neutral fiber and the axial forces are balanced ($N_\alpha=0$, $\forall \alpha$). Among all possible deformations which transform a beam into a circle, the  one allowing the existence of the neutral fiber costs the minimum amount of work. {\bf B:} The undeformed condition is a circle. The line of centroids ($\alpha=1/2$) is depicted as a dashed red circle, while the material fiber granting the axial-bending uncoupling, defined by $\alpha_U$ in  Eq.~\eqref{alpha_NF}, is shown as a dotted black line. {\bf B$'$:} The circle in panel B is deformed into a straight beam. If the fiber identified by $\alpha_U$ keeps its length constant during the circle extension, therefore it coincides with the neutral fiber and the Parent's principle is satisfied. Any other stretching of the circle into a  bar will cost more energy than that leaving unchanged the neutral fiber. Notice that, although the circles in panels A$'$ and B have the same dimensions, the absence of axial forces  implies that the beam in A is longer than that in B$'$ (see SI). {\bf C:} The generic undeformed configuration has a line of centroids shown as a dashed red line. The curve allowing the axial-bending uncoupling in the Eq.\eqref{E_ct} is show as black dotted line. This curve does not overlap with any material fiber according to \eqref{alpha_NF_ct} {\bf C$'$:} Any transformation where the length of the black dotted line is left invariant fulfills the Parent's principle. Hence, the notion of neutral fiber shifts to that of \emph{neutral curve}. Here the final deformed condition if flat, showing how the neutral curve is deformed differently from any other material fiber. }
\label{fig2}
\end{figure}

The neutral fiber is the locus of points which does not undergo any longitudinal extension nor contraction during the elastica deformation~\cite{love2013treatise}. If there is one, this does not necessarily coincide with the  line of centroids~\cite{love2013treatise} and its existence depends entirely on the transformation put in use.

\noindent After the Jacob Bernoulli's first erroneous attempt of establishing a general principle for the neutral fiber placement, based on the balance of bending moments between elongated and compressed
fibers, the correct condition was enunciated by Parent in 1713~\cite{parent1713veritable} and rediscovered by Coulomb 60 years later~\cite{coulomb1773essay}: \emph{the neutral fiber is the locus of points separating the region of extension from that of compression, such that the longitudinal forces are balanced}. This criterion, however, is easy to implement in case of Hookean dependence of the tension over the cross-section, but is hard to be assessed for a generic (non-linear) stress-strain relation, as precisely sought by Jacob Bernoulli.

  Let us see how  Parent's principle applies to the case under investigation. We start by considering an elastica with uniform spontaneous curvature, a situation encompassing the cases of a slender bar, a circular arc or an hyperbola. The requirement of the longitudinal forces balance corresponds to $N_\alpha=0$. However this  constitutes just the necessary condition for the existence of the neutral line. As a matter of fact it does not furnish a precise answer about  its placement because, in view of the first of Eq.~\eqref{const_eq-any_transf}, it does not depend on the specific choice of the representative fiber. 
 However, among all the representative fibers, for rigid cross sections, only one satisfies the zero strain condition $\varepsilon_\alpha=0$, i.e. the value of $\alpha_U$ in~\eqref{alpha_NF}. Therefore, as a general principle, we can conclude that the neutral fiber, when it exists, corresponds to the material representative line which ensures the axial-bending uncoupling of the strain energy. This is an universal property, valid  for  both flat as well as for non-flat initial configurations.  Moreover, it appears that for any transformation guaranteeing the existence of the neutral fiber, the bending moment is independent of the representative fiber choice, as implied by Eq.s~\eqref{const_eq-any_transf}. 

 We now investigate how the neutral fiber is identified when the curvature is not constant throughout the elastica in the undeformed configuration. This is achieved by using the macroscopic constitutive relations in the local formulation~\eqref{const_eq_ct_any}. The condition implied by the Parent principle requires $N_\alpha(s)=0$ $\forall s$, accompanied by the zero strain condition $\left|\mathbf{t_\alpha}(s)\right|=\left|\mathbf{T_\alpha}(s)\right|$.  While in the case of elastica endowed with constant spontaneous curvature the neutral fiber is identified with a material line corresponding to a constant $\alpha$,  the situation is totally different for  a generic undeformed  configuration. As a matter of fact, in this case one cannot properly talk about neutral fiber, in the sense that the curve guaranteeing $N_\alpha(s)=0$  together with the zero local strain conditions, does not overlap with one of the elastica material fibers.   Rather, the concept of neutral fiber is replaced by that of neutral curve, defined  as the locus of points  identified by the value of $\alpha_U(s)$ in Eq.~\eqref{alpha_NF_ct}. A typical situation is that depicted in Fig.~\ref{fig2}C, where the initial curvature is a function of the internal parameter $s$, $K_\alpha(s)$, and the neutral curve is shown as a black dotted line,  jiggling around the line of centroids (dashed red line), and
  approaching the side of the elastica with maximum value of the natural curvature. Notably, while any representative fiber is always orthogonal to the cross-section, the neutral curve is not in general orthogonal.
 
 Therefore, if the elastica transformation  provides the existence of a neutral curve,  it is possible to adopt a \emph{neutral} arc-length parametrization such that $\left|\mathbf{T}_{\alpha_U    }(s)\right|=1$. Adopting this parametrization,  the energy assumes the following simple form (see SI)

\be
\mathscr{E}_{\alpha_U} = -\frac{bYh^2}{2}\int_{s_m}^{s_M}ds\left[\frac{1}{2}-\alpha_U(s)\right]
\,\frac{\left[\varphi'(s)-\Phi'(s)\right]^2}{\Phi'(s)}.
\label{E_ct_NC}
\ee

 \noindent At the same time, the bending moment takes the form

\be
  M_{\alpha_U}(s)=-bYh^2\left[\frac{1}{2}-\alpha_U(s)\right]\,\frac{\left[\varphi'(s)-\Phi'(s)\right]}{\Phi'(s)}.
  \label{bending_mom_ct_NC}
\ee   

\noindent Let us take a moment to clarify what is implicit in the expression~\eqref{E_ct_NC}. The neutral curve requires that the value of $\alpha$ changes with $s$ according to Eq.~\eqref{alpha_NF_ct}. As a consequence, the neutral arc-length parameterization entails that the strain energy~\eqref{E_ct_NC} interpolates the expressions~\eqref{E_ct}, in the same fashion as $\alpha_U(s)$ interpolates among the different $\alpha$. However, for a deformation that preserves the neutral curve, the value of $\mathscr{E}_{\alpha_U} \equiv \mathscr{E}_{\alpha}$, thanks to the property of invariance of $\Delta E_\alpha$ under a change of reference frame, which holds locally also in the continuum limit.

\section{Discussion} \label{sec:disc}

 Our one-dimensional treatment of a three-dimensional elastica is in the wake of the old theories of flexure and bending. Nevertheless, the choice of a fiber  $\alpha$ as the  representative one-dimensional medium is purely arbitrary. As a matter of fact, the linchpin of our analysis has been to disentangle the representative material fiber from the notion of neutral fiber. Indeed, contrary to the modern three-dimensional theories of bending, we do not postulate the existence of a neutral fiber, neither we identify it with the line passing through the centroids. Our main achievement has been to furnish the exact criteria for its existence and its location. These criteria ultimately stem from the derivation of the strain energy, and are encapsulated in the following universal condition:

\begin{equation}
    \left.\frac{\partial \Delta E_\alpha}{\partial\varepsilon_\alpha}\right|_{\varepsilon_\alpha=0}=0.
    \label{NF_final}
\end{equation}

\noindent 
 Locating the position of the neutral fiber does not have only an effect on the stiffness of  a given cross-sectional form, the Eq.~\eqref{NF_final} implies that it becomes a crucial issue for any physical transformation aiming at minimizing the amount of work required, given a shape transformation. Let us explain this point with the help of  Fig.~\ref{fig2}: imagine to have a slender bar, such that in panel A, transformed into the circle in panel A$'$. Among all the possible deformations that bend the elastic bar  into a ring, only one allows the presence of the neutral fiber, and is that leaving unvaried the line of the centroids. In this case,  Parent's principle is satisfied and the longitudinal forces are balanced, i.e. $N_\alpha=0$ $\forall \alpha$.  We could imagine to give the bar  a final circular shape  preserving  the length of a different fiber, say the bottom material line ($\alpha=0$), together with Jacob Bernoulli (in his first attempt, dating 1694) or Euler \cite{euler1960rational}. In this case, however, $N_\alpha\neq 0$. Therefore the bottom line cannot be identified with the neutral fiber although $\varepsilon_{\alpha=0} = 0$. The important point is that  the amount of work spent by deforming the beam into a circle,  is minimum if the line of centroids maintains its length unvaried. This is plainly demonstrated in  SI.

\noindent Now, it is also instructive to consider the opposite case, where the circular shape is the unstressed condition (Fig.~\ref{fig2}B), and the flat configuration is the  deformed one (Fig.~\ref{fig2}B$'$). The neutral fiber in this case does not correspond to the line of centroids (red dashed line) but it is a circumference whose radius is  smaller, according to the value of~\eqref{alpha_NF} (black dotted line). Thus the minimum amount of work required to ``stretch'' the circle into a bar is obtained if the corresponding transformation leaves the neutral fiber unvaried. As a corollary, the bar shown in panel B$'$ of Fig.~\ref{fig2} is smaller than that in panel A, because the length of the extended beam in panel B$'$ is $l=\frac{2\pi h}{\ln\left(\frac{L+\pi h}{L-\pi h}\right)}<L$. 

So far we addressed the cases  of elastica endowed with spontaneous constant curvature. We now turn to the elastica with intrinsic local curvature as in Fig.~\ref{fig2}C.    The variational version of the minimum principle in~\eqref{NF_final} is 
\begin{equation}
\left.\frac{\delta \mathscr{E}_\alpha}{\delta \left|\mathbf{t_\alpha}(s)\right|}\right|_{\left|\mathbf{t_\alpha}(s)\right|=\left|\mathbf{T_\alpha}(s)\right|}=0.
    \label{NF_final_ct}
\end{equation}
\noindent Imagine an elastic filament undergoing a shape transformation. One is generally inclined to think that such a system would prefer to deform by minimizing its  energy,  skewing and swelling in the fashion that is the easiest of all. The expression~\eqref{NF_final_ct} specifies  that such a minimum-cost deformation is only obtainable by preserving the length of the neutral curve. However, due to the immaterial nature of the  neutral curve, its profile  may differ considerably from the actual three-dimensional  configuration. This is graphically represented in Fig.~\ref{fig2}C$'$ where, for the sake of simplicity, the elastica deformed configuration is set to be straight and  the neutral curve is quite something else. 

\subsection{The importance of the neutral curve for ideal chains}

The above considerations have a tremendous impact into both non-equilibrium properties and equilibrium statistics of polymers or other macromolecules. Take for example the conformational dynamics of ideal chains~\cite{rubinstein2003polymer,doi1988theory}. Ideal chains provide  simplified one-dimensional models for real flexible and semiflexible biomolecules, being substantially the trace of their  backbone, with  negligible self-interactions
or excluded volume effects.    Among  semiflexible models,  certainly the most successful is the Worm-Like Chain (WLC) introduced by Kratky and Porod
\cite{kratky1949rontgenuntersuchung,marko1995stretching} for inextensible stiff-rod polymers, for which  the energy associated with conformational fluctuations may be captured by using merely linear elasticity. The effective WLC energy associated with the bending is ~\cite{fixman1973polymer,yamakawa1977statistical,marko1995stretching}

\begin{equation}
    \mathscr{E}_{WLC}=   \frac{k_BT}{2}\int_0^Lds\,A\,k(s)^2
    \label{E_WLC}
\end{equation}

\noindent where $k_B$ is the Boltzman constant, $T$ is the temperature and $A$ is the persistent length, characterizing the bending stiffness of the polymer. As the expression~\eqref{E_WLC} is derived from thin-rod linear elasticity, it entirely fits into the framework developed in this paper. Indeed the stretching energy of a straight ribbon with inextensible neutral fiber, i.e. the line of centroids, can be expressed as    

\begin{equation}
    \mathscr{E}_{1/2}=   \frac{bYh^3}{2}\int_0^Lds\,\frac{\varphi'(s)^2}{12}
    \label{E_ct_flat_1/2}
\end{equation}

\noindent if the same line of centroids ($\alpha=1/2$) is assumed as the representative fiber (see SI). The analogy between the expression~(\ref{E_WLC}) and~\eqref{E_ct_flat_1/2} is apparent if one recalls that $i)$ both expressions are derived assuming the arc-length parametrization for which $\left|\mathbf{t}_{(1/2)}(s)\right|=1$, $ii)$ the local curvature $k(s)\equiv \left|\varphi'(s)\right|$, and $iii)$  the Young's modulus $Y$ exhibits a temperature dependence~\cite{varshni1970temperature} (in this respect notice that also $A$ may depend on $T$~\cite{martin2018temperature,geggier2011temperature}). At the same time, from the comparison of Eq.s~\eqref{E_WLC} and~\eqref{E_ct_flat_1/2} it  emerges that the natural configuration of a polymer is tacitly assumed to be a rigid rod~\cite{fuller1971writhing, tanaka1985elastic}, not only at $T=0$. This contrasts with real polymeric chains, whose microstructural shape is naturally coiled.  Taking into account the polymers' natural curvature requires the formal extension of the ideal energy~\eqref{E_WLC} along the line marked by our theoretical analysis. In particular, in view of Eq.~\eqref{NF_final_ct}, the inextensibility requirement attains the higher value of a principle of minimum energy. To be inextensible, however, is none of the polymer material lines, but the neutral curve. This practically translates into fluctuations of the three dimensional polymer conformation and of its contour length, although limited, in contrast with the classical WLC picture. Importantly, it  elucidates the contour molecule fluctuations observed in short  to moderate length molecules, where $L/A \sim 6-20$~\cite{seol2007elasticity}, and it may help to explain the extreme bendability of short DNA~\cite{vafabakhsh2012extreme}. As a matter of fact, the difference between the length of a generic fiber and that of the natural curve becomes more and more apparent for short filaments, or for increasing values of the $h/L$ ratio.  Hence, assuming the neutral arc-length parametrization, the correct expression for the bending energy is the Eq.~\eqref{E_ct_NC}, while any other material fiber representation based on constant $\alpha$ requires the general expression~\eqref{E_ct}. This contrasts with flexible polymers for which the  concept of neutral curve loses importance and, even assuming the neutral arc-length parametrization~\eqref{alpha_NF_ct}, the energy is given by 

\be
\mathscr{E}_{\alpha_U} = \frac{bY}{2}\int_{s_m}^{s_M}ds\, \left\{h \left[\left|\mathbf{t}_{\alpha_U}(s)\right|-1\right]^2-h^2\left[\frac{1}{2}-\alpha_U(s)\right]
\,\frac{\left[\varphi'(s)-\Phi'(s)\right]^2}{\Phi'(s)}\right\}.
\label{E_ct_flex}
\ee

\noindent Finally, from a statistical mechanics point of view, at the equilibrium the Boltzmann distribution is $e^{-\frac{\mathscr{E}_{\alpha_U}}{k_BT}}$ rather than $e^{-\frac{\mathscr{E}_{WLC}}{k_BT}}$, weighting a statistical ensemble composed by polymers configurations satisfying the constrain of having constant neutral curve rather than constant line of centroids.

\subsection{Numerical model for vesicles and polymers}

Models for  polymers, flexible films, membranes or vesicles often require a numerical implementation, in order to 
study shapes, fluctuations and dynamics, complementing  the statistical mechanics analytical approach~\cite{nelson1989statistical}. These  models in general consist of $N$ impenetrable circular beads connected by links, arranged in a configuration which requires  a well defined energy cost. A typical example is furnished by planar thermally fluctuating rings used as simplified
models of fluctuating vesicles such as red blood cells. The Liebler-Singh-Fisher (LSF) model~\cite{leibler1987thermodynamic}, by instance, considers planar closed chains including the pressure and a curvature energy term such as $\sum_{i=1}^N \frac{k}{\Delta L} \left(1-\cos\Delta \varphi^{(i)}\right)$, where $k$ is the constant bending modulus.  Many aspect and variants of the 
LSF have been discussed in the physics 
literature using a wealth of analytical and computational techniques~\cite{levinson1992asphericity,gaspari1993shapes,haleva2006smoothening,mitra2008phase,maggs1990size,camacho1991semiflexible,camacho1990tunable,fisher1989fractal,landau2012computer}, allowing the stretching of the ring~\cite{romero1992simple,marconi1993deflated}, as well as a spontaneous curvature and locally varying bending modulus~\cite{katifori2009collapse}.

In our theory the strain  energy function~\eqref{E_ct} has been derived by a finite difference scheme, as the continuum limit
of the sum of the blocks strain energy~\eqref{E_block}. Therefore, the discrete  form of the Eq.~\eqref{E_ct} constitutes, without any further approximation, the appropriate expression to be used in numerical simulations (see Fig.~\ref{fig1}B,B$'$ and SI). However, the fact that the uncoupling value of $\alpha_U$ in~\eqref{alpha_NF} depends crucially on the local value of the curvature of each block, makes it impossible to adopt a global neutral arc-length parametrization for discrete chains. This means that a discrete decoupled expression for the strain energy function is out of the question, because the discrete chain  would  result  in a discontinuous polygonal, when $N$ is finite (green dashed lines in Fig.~\ref{fig1}B,B$'$). From an energetic point of view, there would be nothing wrong, as the energy would remain the same due to the strain energy's invariance under a change of reference frame. However, from a graphical perspective, it would look awful! On the other side, adopting the constant $\alpha$ description yields the correct and more convenient expression for the numerical implementation of our model, valid for any value of $N$:  

\begin{equation*}
    E_\alpha = \frac{Y}{2}\sum_{i=1}^N\Delta L_\alpha^{(i)}\left\{F^{(i)}_\alpha\, {\varepsilon^{(i)}_\alpha}^2+2\frac{S^{(i)}_\alpha}{\Delta L_\alpha^{(i)}} \varepsilon^{(i)}_\alpha\left[\Delta\Phi^{(i)}-\Delta\varphi^{(i)}\right]+ \right.
\end{equation*}
\be
\left. +\frac{I^{(i)}_\alpha}{\Delta L_\alpha^{(i)2}}\left[\Delta\Phi^{(i)}-\Delta\varphi^{(i)}\right]^2\right\} \ . 
    \label{meth_E_block_dscrt}     
\ee

\noindent We emphasize that this expression holds for any elastic chain, be extensible or unextensible. The unextensibility condition should be enforced by keeping the lengths of the segments at the height $\alpha_U^{(i)}$ constants, in any conformational change, and modifying the representative $\alpha$ fiber accordingly. However, the detailed procedure of the numerical implementation of our model will be the subject of an upcoming publication.   

\subsection{Real filaments}

Real biological filaments are characterized by local properties that can vary smoothly or abruptly along their contour. Our analysis focused on what we believe to be the most prominent property: the spontaneous curvature. However, geometric and mechanical properties may also characterize different portions of cellular filaments. In reality, filaments can only be considered homogeneous on average, with different heights ($h$) and depths ($b$) identifying each discrete protein-based segment. Simultaneously, the intrinsic mechanical properties of each monomer can differ, as in the case of the elastic modulus ($Y$), which we assumed to be constant in our analysis.

 We believe that the entire framework assembled in this paper can be adapted to handle and accurately describe realistic biological filaments. The reasons lie in two main ingredients of our theory: the discrete scheme and the strain energy invariance under a change of reference frame. The finite difference scheme allows for a more precise description of biological shapes than any continuum theory, such as Cosserats' theory~\cite{cosserat1909theorie,rubin2000cosserat,naghdi1984constrained,green1974theory1,green1974theory2} because any biological filament, whether it is a microtubule, actin, or intermediate filament, is ultimately composed of discrete units. Since the blocks are independent in our model, the geometrical and mechanical characteristics of biological filaments can be easily implemented locally, while preserving the general form of the strain energy as in \eqref{E_block_dscrt}. 

Furthermore, while the concept of material fibers becomes meaningless for non-homogeneous filaments, or at least non-parallel ones, the parametrization in terms of a reference segment $\alpha\in[0,1]$, inherent to any block, will always be possible, giving the concept of fiber a geometrical rather than material meaning. Most importantly, the existence of a neutral curve, allowing the axial-bending decoupled form of the strain energy, must remain valid in any context, highlighting once more the general value of our findings.

\section*{Acknowledgements} \label{sec:acknowledgements}
We thank for the useful discussions, criticisms and suggestions Stefano Zapperi, Lev Truskinovsky , Silas Alben, Umut Orguz Salman and Daniel Rayneau-Kirkhope.

 \bibliographystyle{elsarticle-num} 
 \bibliography{elasticity}






\end{document}


\title{General theory for plane extensible elastica with arbitrary undeformed shape \\ SUPPLEMENTARY INFORMATION}

\author{Alessandro Taloni*}
    \email[Corresponding author; ]{alessandro.taloni@isc.cnr.it}
    \affiliation{CNR -- Consiglio Nazionale delle Ricerche, Istituto dei Sistemi Complessi, via dei Taurini 19, 00185 Roma}
    \affiliation{Center for Complexity and Biosystems, Department of Physics, University of Milan, Milan, Italy}

\author{Daniele Vilone}
    \email[Correspondence email address: ]{daniele.vilone@gmail.com}
    \affiliation{Laboratory of Agent Based Social Simulation, Institute of Cognitive Science and Technology, National Research Council, Via Palestro 32, 00185 Rome, Italy}
    \affiliation{Grupo Interdisciplinar de Sistemas Complejos, Departamento de Matem\'aticas, Universidad Carlos III de Madrid, 28911 Legan\'es, Spain}
\author{Giuseppe Ruta}
    \email[Correspondence email address: ]{giuseppe.ruta@uniroma1.it }
    \affiliation{Dipartimento di Ingegneria strutturale e geotecnica, Sapienza Università di Roma. Via Eudossiana n° 18, 00184 Roma}


\



\maketitle

\input{methods}

\bibliographystyle{unsrt}
\bibliography{elasticity}

%% file: methods.tex
\section{Derivation of the strain energy from a generic undeformed configuration \label{sec:meth_derivation_any}}

\begin{figure}[h]
\centering
\includegraphics[width=9.1cm]{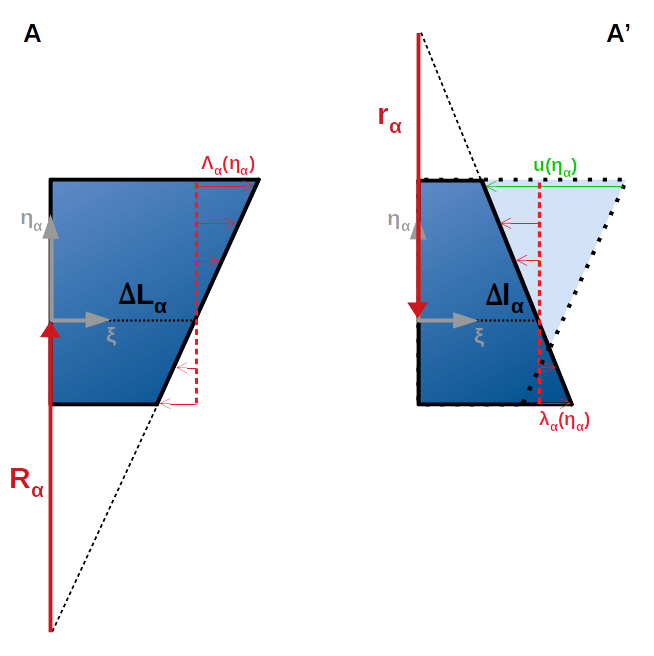}
\caption{Strain energy from a generic undeformed configuration. {\bf A}: The block  undeformed configuration is such that the representative material line has a natural extension equal to $\Delta L_\alpha$. Correspondingly, the size of any other fiber is expressible through the relation \eqref{meth_elong_any}, where the quantity $\Lambda(\eta_\alpha)$
are represented by orizontal thin red arrows. The spontaneous 
radius of curvature relative to the representative fiber $R_\alpha$ is shown by a vertical thick red arrow. {\bf A $'$}: In the deformed  condition, the size of the representative fiber varies from $\Delta L_\alpha$ to $\Delta l_\alpha$ and its radius of curvature becomes $r_\alpha$. In the reference frame integral with the block, any fiber undergoes a displacement $u(\eta_\alpha)$ (green arrows). At the same time, the size of any deformed  material fiber can be expressed as $ \Delta l(\eta_\alpha)= \Delta l_\alpha + \lambda_\alpha(\eta_\alpha)$ (red arrows).}
\label{figS5}
\end{figure}

We consider the elementary block depicted in Fig.~\ref{figS5}. The undeformed condition is represented in panel A, whereas its deformed state is shown in panel A$'$. The length of the representative fiber in the natural state is $\Delta L_\alpha$, and that of a generic fiber at an height $\eta_\alpha$ is

\be
 \Delta L(\eta_\alpha)= \Delta L_\alpha + \Lambda_\alpha(\eta_\alpha)
 \label{meth_elong_any}.
 \ee
 
 \noi The corresponding deformed quantities are  $\Delta l_\alpha$ and $\Delta l(\eta_\alpha)$ are defined by

 \be
 \Delta l(\eta_\alpha)= \Delta l_\alpha + \lambda_\alpha(\eta_\alpha)
 \label{meth_elong_1_flat}.
 \ee

 \noi Hence, the deformation that a generic material segment  experiences  passing from the state A to the state A$'$ is expressed as

  \be
 u(\eta_\alpha)= \Delta l_\alpha - \Delta L_\alpha + \Lambda_\alpha(\eta_\alpha) - \lambda_\alpha(\eta_\alpha) 
 \label{meth_elong_1_any}.
 \ee

\noi The geometry of the undeformed shape satisfies the following equality \cite{euler1960rational,timoshenko1983history}

\be
\frac{\Delta L_\alpha}{R_\alpha}=\frac{\Lambda_\alpha(\eta_\alpha)}{\eta_\alpha},
 \label{meth_geom_rel_undef}
\ee

\noi while for the deformed configuration we have

\be
\frac{\Delta l_\alpha}{r_\alpha}= \frac{\lambda_\alpha(\eta_\alpha)}{\eta_\alpha}\ .
    \label{meth_geom_rel}
\ee

\noi  Thanks to Eqs.~(\ref{meth_geom_rel_undef}) and~(\ref{meth_geom_rel}), the displacement~(\ref{meth_elong_1_any}) attains the final form  

 \be
 u(\eta_\alpha)= \Delta l_\alpha - \Delta L_\alpha + \eta_\alpha\left( \frac{\Delta L_\alpha}{R_\alpha} - \frac{\Delta l_\alpha}{r_\alpha}\right) 
 \label{meth_elong_any_final}.
 \ee
 
\noindent By substitution of Eqs.~(\ref{meth_elong_any}) and~(\ref{meth_elong_any_final}) into

\be
\Delta E_\alpha=\frac{bY}{2}\int_{-\alpha h}^{(1-\alpha)h}\left[\frac{u(\eta_\alpha)}{\Delta L(\eta_\alpha)}\right]^2 \Delta L (\eta_\alpha)\ d\eta_\alpha , 
    \label{meth_E_block}
\ee

\noi  we obtain the expression:

\be
\Delta E_\alpha = \frac{Y}{2\,\Delta L_\alpha}\left[
F_\alpha\left(\Delta l_\alpha - \Delta L_\alpha\right)^2
+
2S_\alpha\left(\Delta l_\alpha - \Delta L_\alpha\right)\left(\frac{\Delta l_\alpha}{r_\alpha}-\frac{\Delta L_\alpha}{R_\alpha}\right)
+I_\alpha\left(\frac{\Delta l_\alpha}{r_\alpha}-\frac{\Delta L_\alpha}{R_\alpha}\right)^2
\right],
\label{meth_E_block_any}
\ee

\noi where

\be
F_\alpha=\int_{-\alpha h}^{(1-\alpha)h}d\eta_\alpha \frac{b}{1+\frac{\eta_\alpha}{R_\alpha}}
\label{meth_F_alpha},
\ee

\be
S_\alpha=\int_{-\alpha h}^{(1-\alpha)h}d\eta_\alpha \frac{b\eta_\alpha}{1+\frac{\eta_\alpha}{R_\alpha}}
\label{meth_S_alpha}
\ee

\noi and

\be
I_\alpha=\int_{-\alpha h}^{(1-\alpha)h}d\eta_\alpha \frac{b\eta_\alpha^2}{1+\frac{\eta_\alpha}{R_\alpha}}
\label{meth_I_alpha}
\ee

\begin{figure}[h]
\centering
\includegraphics[width=10.1cm]{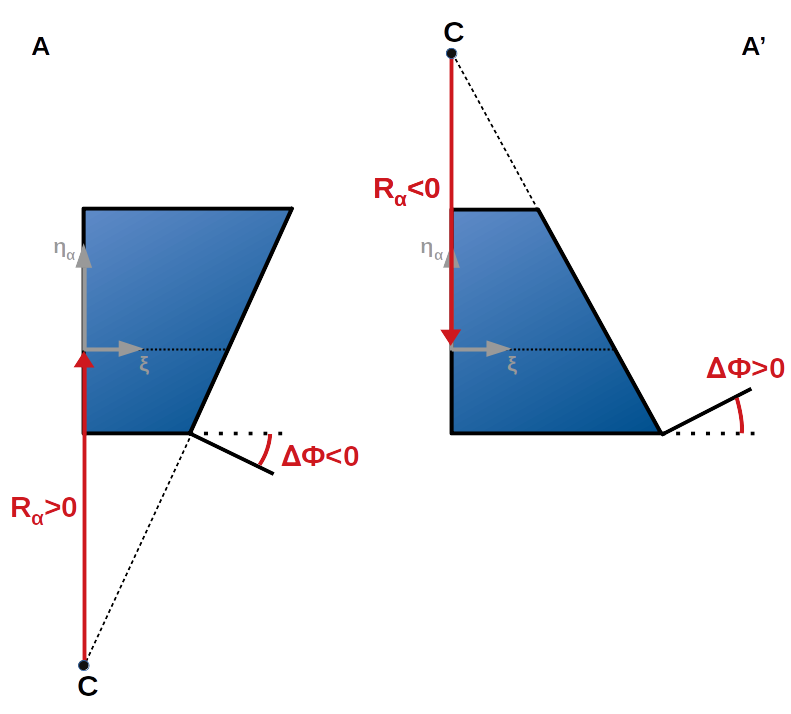}
\caption{Sign of the radius of curvature. The radius of curvature has a sign assigned whether its direction, connecting the intersection $C$ of the sidelines containing the block's sections with the beginning of the reference segment, coincides with that of the $\eta_\alpha$ axis of the frame integral with the block (panel {\bf A}), or opposite to it (panel {\bf A$'$}). }
\label{figS6}
\end{figure}

\noi These three factors have different functional forms according to whether $R_\alpha$ is positive or negative (see Fig.~\ref{figS6}). The quantity  $R_\alpha$ can be positive or negative, as required by the consistency of the Eq.(\ref{meth_geom_rel_undef}). $\left|R_\alpha\right|$ becomes the radius of the osculating circle which locally approximates the reference segment in the continuum limit, and the  sign of $R_\alpha$ is assigned in the following way. It is clear that the intersection \emph{C} between the sidelines containing the block's sections lies on the axis $\xi=0$ of the local reference system $\xi\text{-}0\text{-}\eta_\alpha$.
If \emph{C} lies below the bottom fiber $\alpha=0$, then $R_\alpha$ is positive (Fig.\ref{figS6}A). If conversely \emph{C} is above the upper fiber $\alpha=1$, then $R_\alpha$ is negative (Fig.\ref{figS6}A$'$).
It follows that \emph{C} has coordinates $(0,-R_\alpha)$ in the local reference system. The same prescription for the sign applies to $r_\alpha$. If $R_\alpha>0$ the solutions of~(\ref{meth_F_alpha}), (\ref{meth_S_alpha}) and~(\ref{meth_I_alpha}) read

\be
F_\alpha=bR_\alpha\ln\left(1+\frac{h}{R_0}\right)
\label{meth_F_alpha>0},
\ee

\be
S_\alpha=bR_\alpha\left[h-R_\alpha\ln\left(1+\frac{h}{R_0}\right)\right]
\label{meth_S_alpha>0}
\ee

\noi and

\be
I_\alpha=bR_\alpha\left[\left(\frac{1}{2}-\alpha\right)h^2-hR_\alpha+R_\alpha^2\ln\left(1+\frac{h}{R_0}\right)\right],
\label{meth_I_alpha>0}
\ee

\noi with $R_\alpha=R_0+\alpha h$. When we consider the case $R_\alpha<0$, the three integrals \ref{meth_S_alpha}-\ref{meth_I_alpha} can be solved yielding

\be
F_\alpha=-bR_\alpha\ln\left(1-\frac{h}{R_1}\right)
\label{meth_F_alpha<0},
\ee

\be
S_\alpha=bR_\alpha\left[h+R_\alpha\ln\left(1-\frac{h}{R_1}\right)\right]
\label{meth_S_alpha<0}
\ee

\noi and

\be
I_\alpha=bR_\alpha\left[\left(\frac{1}{2}-\alpha\right)h^2-hR_\alpha-R_\alpha^2\ln\left(1-\frac{h}{R_1}\right)\right],
\label{meth_I_alpha<0}
\ee

\noi where  $R_\alpha=R_1-(1-\alpha)h$. The two opposite bending states in Fig.\ref{figS6} are expressible in the  compact forms provided in the main text recalling that, if $R_\alpha>0$, hence necessarily $R_0<R_1$ ($K_0>K_1$), whilst, for $R_\alpha<0$, therefore $|R_1|<|R_0|$ and $K_1>K_0$. We stress the fact that the expressions  of $F_\alpha$, $S_\alpha$ and $I_\alpha$  are fully established once the value of $\alpha$ and $\max[K_0,K_1]$ are furnished. In particular  the following identity holds if $K_0>K_1$

\be
K_\alpha = \frac{K_0}{1+\alpha h K_0},
\label{Kalpha_definition_+}
\ee

\noi while for $K_1>K_0$

\be
K_\alpha =  \frac{K_1}{1+(1-\alpha) h K_1}.
\label{Kalpha_definition_-}
\ee

\begin{figure}[h]
\centering
\includegraphics[width=12.1cm]{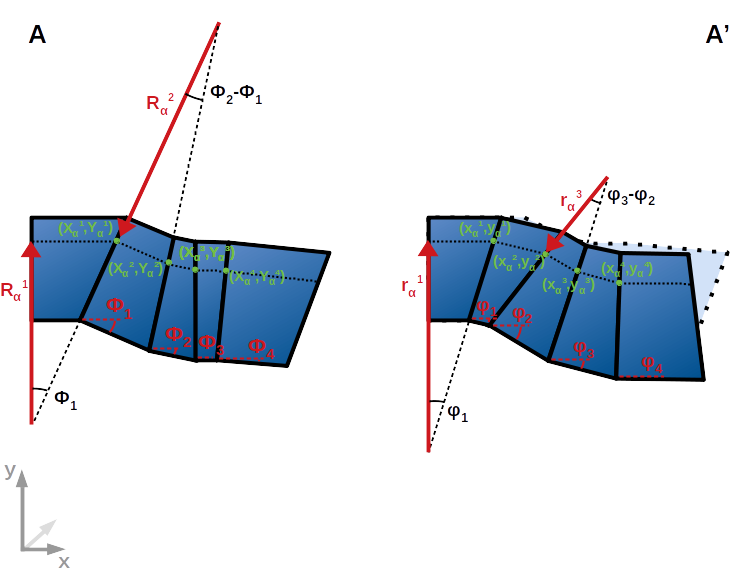}
\caption{Beam discrete strain energy from a generic undeformed configuration. {\bf A}: The polygonal, connecting the vertices of each undeformed representative segments, is represented by a dotted black thick line.
In the laboratory frame, the vertices have planar coordinates $\left(X_\alpha^{(i)},Y_\alpha^{(i)}\right)$ shown as green dots. The spontaneous radius of curvature can be positive (as $R_\alpha^1$) or negative (as $R_\alpha^2$). The connection between  the size of representative segment, $\Delta L_\alpha^{(i)}$, and  its spontaneous radius of curvature $R_\alpha^{(i)}$, is given by the relation \eqref{meth_bending_angle_undef}, where the bending angles $\Phi^{(i)}$ define the deflection of the undeformed blocks  from the external $x$ axis. {\bf A$'$}: each block  composing the beam undergoes a deformation, such that the deformed polygonal chains  $l_\alpha$ is constructed by the sequence of the representative segments whose vertices are $\left(x_\alpha^{(i)},y_\alpha^{(i)}\right)$ (green dots).}
\label{figS7}
\end{figure}

\noi According to the integral expressions \eqref{meth_F_alpha} and \eqref{meth_I_alpha}, it is always $F_\alpha>0$ and $I_\alpha>0$ because the integrand functions are strictly positive in the integration interval. To the contrary, the sign of $S_\alpha$ can vary according to the value of $\alpha$ and to  $\max[K_0,K_1]$ ($\min[R_0,R_1]$). For the sake of clarity,   $S_{\alpha}>0$ for $\alpha<\alpha_U$, and $S_{\alpha}<0$ for $\alpha>\alpha_U$. 
The choice of $\alpha_U$  ensuring the stretching-bending uncoupling, guarantees that

\be
F_{\alpha_U}=bh
\label{meth_F_alpha_decoupling},
\ee

\be
S_{\alpha_U}=0
\label{meth_S_alpha_decoupling}
\ee

\noi and

\be
I_{\alpha_U}=\frac{b}{K_{\alpha_U}}\left(\frac{1}{2}-\alpha_U\right)h^2=\frac{\mbox{sgn}\left(K_0-K_1\right)bh^3}{\ln\left(1+h\,\max[K_0,K_1]\right)}\left[\frac{1}{2}-\frac{1}{\ln\left(1+h\,\max[K_0,K_1]\right)}+\frac{1}{h\,\max[K_0,K_1]}\right]
\label{meth_I_alpha_decoupling}
\ee

The finite difference scheme, outlined so far, requires  the evaluation of the discrete strain energy $E_\alpha=\sum_{i=1}^N \Delta E_\alpha^{(i)}$ needed to deform the elastica in Fig.~\ref{figS7} from A to A$'$. Moving to the laboratory frame we find that the relation

\be
\frac{\Delta L_\alpha^{(i)}}{R_\alpha^{(i)}}=-\tan\Delta \Phi^{(i)}
\label{meth_bending_angle_undef}.
\ee

\noi is always satisfied \cite{greenberg1967equilibrium}, where $\Delta \Phi^{(i)}=\Phi^{(i)}-\Phi^{(i-1)}$ and $\Phi^{(i)}$  is the $i$-th cross-sectional bending angle with respect the $x$ axis. Since we assume the limit of small deflections, we can approximate $\tan\Delta \Phi^{(i)}\simeq\Delta \Phi^{(i)}$. 
The reference undeformed polygonal chain $L_\alpha$ is specified by the series of points $\mathbf{L}_\alpha^{(i)}=(X_\alpha^{(i)},Y_\alpha^{(i)})$, with line segments $\Delta L_\alpha^{(i)}=\left|\mathbf{L}_\alpha^{(i)}-\mathbf{L}_\alpha^{(i-1)}\right|$ (Fig.\ref{figS7}A).

\noi In the deformed state (Fig.\ref{figS7}A$'$) we have

\be
\frac{\Delta l_\alpha^{(i)}}{r_\alpha^{(i)}}=-\tan \Delta \varphi^{(i)}
\label{meth_bending_angle}
\ee

\noi where $\Delta \varphi^{(i)}=\varphi^{(i)}-\varphi^{(i-1)}$, and $\varphi^{(i)}$ corresponds to the bending angle between the $i-$th block and  the $x$ axis. Again we assume the small deflection limit, i.e. $\tan \Delta \varphi^{(i)}\simeq \Delta \varphi^{(i)}$. The deformed chain $l_\alpha$ has the points $\mathbf{l}_\alpha^{(i)}=(x_\alpha^{(i)},y_\alpha^{(i)})$ as vertices,  with $\Delta l_\alpha^{(i)}=\left|\mathbf{l}_\alpha^{(i)}-\mathbf{l}_\alpha^{(i-1)}\right|$. We introduce the strain measure as $\varepsilon_\alpha^{(i)}=\frac{\Delta l_\alpha^{(i)}-\Delta L_\alpha^{(i)}}{\Delta L_\alpha^{(i)}}$, while the bending strain measure can be obtained by two different definitions: the first is due to Kammel \cite{kammel1966einfluss}

\be
\mu_\alpha^{(i)}=\frac{\Delta \Phi^{(i)}-\Delta \varphi^{(i)}}{\Delta L_\alpha^{(i)}}
\label{meth_bendging_meas_def3},
\ee

\noi and the second to Antman \cite{antman1968general}

\be
\mu_\alpha^{(i)}=\frac{\Delta l_\alpha}{\Delta L_\alpha}\frac{1}{r_\alpha}-\frac{1}{R_\alpha}
\label{meth_bendging_meas_def4}.
\ee

\noi One can easily see that they are equivalent by the construction in Fig.\ref{figS3}. Upon summation of the terms in Eq.~(\ref{meth_E_block_any}), using the definition \eqref{meth_bendging_meas_def4}, we obtain the  energy

\be
E_\alpha = \frac{Y}{2}\sum_{i=1}^N\Delta L_\alpha^{(i)}\left\{F^{(i)}_\alpha\, {\varepsilon^{(i)}_\alpha}^2+2S^{(i)}_\alpha\, \varepsilon^{(i)}_\alpha\mu^{(i)}_\alpha+I^{(i)}_\alpha\, {\mu^{(i)}_\alpha}^2\right\} \ . 
    \label{meth_E_block_dscrt}     
\ee

\noi The expression of $F_\alpha$, $S_\alpha$ and $I_\alpha$ in terms of $\Delta\Phi$, $\Delta L_\alpha$ and $\alpha$, is achieved by inserting the relation~(\ref{meth_bending_angle_undef}) into the expressions~(\ref{meth_F_alpha>0})--(\ref{meth_I_alpha>0}) and~(\ref{meth_F_alpha<0})--(\ref{meth_I_alpha<0}):
\be
F_\alpha^{(i)} = \frac{\Delta L_\alpha^{(i)}}{\left|\Delta\Phi^{(i)}\right|}\ln\left(1+h\frac{\left|\Delta\Phi^{(i)}\right|}{\min[\Delta L_0^{(i)},\Delta L_1^{(i)}]}\right),
\label{meth_F_alpha_discrt}
\ee

\be
S_\alpha^{(i)} = -\frac{\Delta L_\alpha^{(i)}}{\Delta\Phi^{(i)}}\left(h-F_\alpha^{(i)}\right)
\label{meth_S_alpha_discrt}
\ee

\noi and

\be
I_\alpha^{(i)} = -\frac{\Delta L_\alpha^{(i)}}{\Delta\Phi^{(i)}}\left[\left(\frac{1}{2}-\alpha\right)h^2-S_\alpha^{(i)}\right]
\label{meth_I_alpha_discrt}.
\ee

\noi We recall that it is convenient to take $\Delta L_\alpha=\Delta L_0-\alpha h\Delta \Phi$ if $\min[\Delta L_0,\Delta L_1]=\Delta L_0$, and $\Delta L_\alpha=\Delta L_1+(1-\alpha) h\Delta \Phi$ if $\min[\Delta L_0,\Delta L_1]=\Delta L_1$. 

The differential strain energy $\mathscr{E}_\alpha$ is derived by firstly introducing two parametric expressions for the undeformed and deformed reference material curves as  $\mathscr{L}_\alpha:[s_m,s_{M}]\to \mathbbm{R}^2$ and  $\ell_\alpha:[s_m,s_{M}]\to \mathbbm{R}^2$ respectively. Secondly we take an arbitrary partition $s_m=s_0<s_1<s_2<\cdots<s_N=s_M$ to which we connect the two polygonal chains $L_\alpha$ and $l_a$, in such a way that the vertices satisfy $\mathbf{L}_\alpha(s_i)\equiv\mathbf{L}_\alpha^{(i)}$ and $\mathbf{l}_\alpha(s_i)\equiv\mathbf{l}_\alpha^{(i)}$. Moreover we define the applications $\Phi:[s_m,s_{M}]\to \mathbbm{R}$ and  $\varphi:[s_m,s_{M}]\to \mathbbm{R}$ with the properties $\Phi(s_i)\equiv\Phi^{(i)}$ and $\varphi(s_i)\equiv\varphi^{(i)}$.
From the definitions of the two strain measures $\varepsilon_\alpha^{(i)}$ and $\mu_\alpha^{(i)}$, it follows

\be
\varepsilon_\alpha(s_i)=\frac{\frac{\left|\Delta \mathbf{l}_\alpha(s_i)\right|}{\Delta s_i} - \frac{\left|\Delta \mathbf{L}_\alpha(s_i)\right|}{\Delta s_i}}{\frac{\left|\Delta \mathbf{L}_\alpha(s_i)\right|}{\Delta s_i}}
\label{meth_semidscrt_strain-any}.
\ee

\noi 

\be
\mu_\alpha(s_i)=\frac{\frac{\Delta \Phi(s_i)}{\Delta s_i} - \frac{\Delta \varphi(s_i)}{\Delta s_i}}{\frac{\left|\Delta \mathbf{L}_\alpha(s_i)\right|}{\Delta s_i}}
\label{meth_semidscrt_bend-any}.
\ee

\noindent In the continuum limit, $N$ is increased until the lengths of the polygonal chains $L_\alpha$ and $l_\alpha$ equal those  of the curves $\mathscr{L}_\alpha$ and $\ell_\alpha$. This condition is mathematically enforced by the limiting relations

\be
\frac{\left|\Delta \mathbf{L}_\alpha(s_i)\right|}{\Delta s_i}\to \left|\mathscr{L}_\alpha'(s)\right|, \,    \ \frac{\left|\Delta \mathbf{l}_\alpha(s_i)\right|}{\Delta s_i}\to \left|\ell_\alpha'(s)\right|
\label{meth_derivative_any}
\ee

\noi as $\Delta s_i\to 0$. Correspondingly, the two tangents to the curves are defined as
$\mathbf{T}_\alpha(s)=\frac{d \mathbf{L}_\alpha}{ds}$ and $\mathbf{t}_\alpha(s)=\frac{d \mathbf{l}_\alpha}{ds}$. Yet, the limit  $\Delta s_i\to 0$ entails

\be
\frac{\Delta \Phi(s_i)}{\Delta s_i}\to \Phi'(s), \,    \ \frac{\Delta \varphi(s_i)}{\Delta s_i}\to \varphi'(s).
\label{meth_derivative_bend_any}
\ee

\noi The differential strain measures follow from the limit of Eq.s \eqref{meth_semidscrt_strain-any} and \eqref{meth_semidscrt_bend-any}:

\be
\varepsilon_\alpha(s)=\frac{ \left|\mathbf{t}_\alpha(s)\right|- \left|\mathbf{T}_\alpha(s)\right|}{\left|\mathbf{T}_\alpha(s)\right|}
\label{meth_ct_strain-any}.
\ee

\noi 

\be
\mu_\alpha(s)=\frac{ \Phi'(s)- \varphi'(s)}{\left|\mathbf{T}_\alpha(s)\right|}
\label{meth_ct_bend-any}.
\ee

\noi Now, if  $\sum_{i=1}^N \Delta L\to \int_0^L ds$, plugging the definitions of $\mathbf{T}_\alpha(s)$, \eqref{meth_ct_strain-any} and  \eqref{meth_ct_bend-any} into Eq.\eqref{meth_E_block_dscrt} we recover the energy reported in the main text:

\be
\mathscr{E}_\alpha = \frac{Y}{2}\int_{s_m}^{s_M}ds
\left\{\frac{F_\alpha(s)}{\left|\mathbf{T_\alpha}(s)\right|}\, \left[\left|\mathbf{t_\alpha}(s)\right|-\left|\mathbf{T_\alpha}(s)\right|\right]^2  - \frac{2S_\alpha(s)}{\left|\mathbf{T_\alpha}(s)\right|}\, \left[\left|\mathbf{t_\alpha}(s)\right|-\left|\mathbf{T_\alpha}(s)\right|\right]\left[\varphi'(s)-\Phi'(s)\right] +
\frac{I_\alpha(s)}{\left|\mathbf{T_\alpha}(s)\right|}\,\left[\varphi'(s)-\Phi'(s)\right]^2\right\},
\label{meth_E_ct}
\ee

\ 

\noindent The differential formula for the three factors $F_\alpha(s)$, $S_\alpha(s)$ and $I_\alpha(s)$ are obtainable from the Eq.s    \eqref{meth_F_alpha_discrt}, \eqref{meth_S_alpha_discrt} and~\eqref{meth_I_alpha_discrt}:

\be
F_\alpha(s) = \left|\frac{\mathbf{T}_\alpha(s)}{\Phi'(s)}\right|\ln\left(1+h\frac{\left|\Phi'(s)\right|}{\min[\left|\mathbf{T}_0(s)\right|,\left|\mathbf{T}_1(s)\right|]}\right),
\label{meth_F_alpha_ct}
\ee

\be
S_\alpha (s)= -\frac{\left|\mathbf{T}_\alpha(s)\right|}{\Phi'(s)}\left(h-F_\alpha(s)\right)
\label{meth_S_alpha_ct}
\ee

\noi and

\be
I_\alpha (s)= -\frac{\left|\mathbf{T}_\alpha(s)\right|}{\Phi'(s)}\left[\left(\frac{1}{2}-\alpha\right)h^2-S_\alpha(s)\right]
\label{meth_I_alpha_ct}.
\ee

\noi It is clear that, when $\min[\left|\mathbf{T}_0(s)\right|,\left|\mathbf{T}_1(s)\right|]=\left|\mathbf{T}_0(s)\right|$, we can express $\left|\mathbf{T}_\alpha(s)\right|=\left|\mathbf{T}_0(s)\right|-\alpha h\Phi'(s)$; conversely, when $\min[\left|\mathbf{T}_0(s)\right|,\left|\mathbf{T}_1(s)\right|]=\left|\mathbf{T}_1(s)\right|$, then $\left|\mathbf{T}_\alpha(s)\right|=\left|\mathbf{T}_1(s)\right|+(1-\alpha)h\Phi'(s)$. Thus, the functional forms of $F_\alpha(s)$, $S_\alpha(s)$ and $I_\alpha(s)$ in Eqs.\eqref{meth_F_alpha_ct}, \eqref{meth_S_alpha_ct} and \eqref{meth_I_alpha_ct} highlight the local character of these quantities and the fact that they are fully established by the values of $\left(\alpha,\Phi',\min[\left|\mathbf{T}_0\right|,\left|\mathbf{T}_1\right|]\right) $.

\noindent To uncouple the bending and stretching contributions in the continuum energy expression $\mathscr{E}_\alpha$, the conditions $S_\alpha(s)=0$ in \eqref{meth_S_alpha_ct} requires that

\be
\alpha_U(s)=
\frac{\left|\mathbf{T}_0(s)\right|}{h\Phi'(s)}-\frac{\left|\Phi'(s)\right|}{\Phi'(s)}\frac{1}{\ln\left(1+h\frac{\left|\Phi'(s)\right|}{\min[\left|\mathbf{T}_0(s)\right|,\left|\mathbf{T}_1(s)\right|]}\right)},
\label{meth_alpha_NF_ct_1}
\ee

\noindent which is also expressible as

\be
\alpha_U(s)=1-\left[\frac{\left|\Phi'(s)\right|}{\Phi'(s)}\frac{1}{\ln\left(1+h\frac{\left|\Phi'(s)\right|}{\min[\left|\mathbf{T}_0(s)\right|,\left|\mathbf{T}_1(s)\right|]}\right)}-
\frac{\left|\mathbf{T}_1(s)\right|}{h   \Phi'(s)}\right].
\label{meth_alpha_NF_ct_2}
\ee

\noindent The two formulae \eqref{meth_alpha_NF_ct_1}
and \eqref{meth_alpha_NF_ct_2} are equivalent and can be obtained recalling that $\left|\mathbf{T}_1(s)\right|=\left|\mathbf{T}_0(s)\right|- h\Phi'(s)$. Therefore  the uncoupling condition is a  local property of the elastica and, for either choice of $\alpha_U(s)$, the three factors \eqref{meth_F_alpha_ct}-\eqref{meth_I_alpha_ct} reduce to

\be
F_{\alpha_U}(s) = bh,
\label{meth_F_alpha_ct_decoupling}
\ee

\be
S_{\alpha_U} (s)= 0
\label{meth_S_alpha_ct_decoupling}
\ee

\noi and

\begin{align}
I_{\alpha_U} (s)&=-\frac{\left|\mathbf{T}_{\alpha_U}(s)\right|}{\Phi'(s)}\left[\frac{1}{2}-\alpha_U(s)\right]bh^2\\
&= -bh^3 \frac{\left|\Phi'(s)\right|}{\Phi'(s)}\frac{1}{\ln\left(1+h\frac{\left|\Phi'(s)\right|}{\min[\left|\mathbf{T}_0(s)\right|,\left|\mathbf{T}_1(s)\right|]}\right)}\left[\frac{1}{2}-\frac{\left|\mathbf{T}_0(s)\right|}{h\Phi'(s)}+\frac{\left|\Phi'(s)\right|}{\Phi'(s)}\frac{1}{\ln\left(1+h\frac{\left|\Phi'(s)\right|}{\min[\left|\mathbf{T}_0(s)\right|,\left|\mathbf{T}_1(s)\right|]}\right)}\right].\qedhere
\label{meth_I_alpha_ct_decoupling}
\end{align}

 The \emph{neutral} arc-length parametrization requires that 

\be
\left|\mathbf{T}_{\alpha_U}(s)\right|=1.
\label{meth_neutral_arc_length}
\ee

\noi Therefore, adopting this parametrization and the neutral curve as representative of the whole elastica we have that for a generic transformation the energy is expressible as 

\be
\mathscr{E}_{\alpha_U} = \frac{bY}{2}\int_{s_m}^{s_M}ds
\left\{h\, \left[\left|\mathbf{t_{\alpha_U}}(s)\right|-1\right]^2 - \frac{h^2}{\Phi'(s)}\left[\frac{1}{2}-\alpha_U(s)\right]\,\left[\varphi'(s)-\Phi'(s)\right]^2\right\}.
\label{meth_E_ct_neutral_arc_length}
\ee

\section{Energy invariance under change of reference frame\label{sec:meth_E_inv}}

 \begin{figure}[h]
\centering
\includegraphics[width=10.1cm]{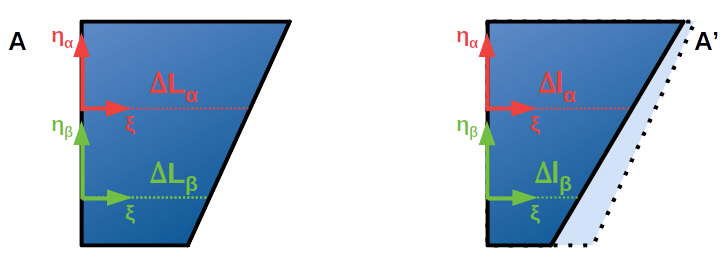}
\caption{Block's energy invariance under change of reference material line.  {\bf A:} In red is represented the reference frame $\xi\text{-}0\text{-}\eta_\alpha$ integral with the block, when the reference material segment is placed at an height $\alpha h$ from the block's bottom surface. The length of the reference segment is $\Delta L_\alpha$. When the reference material segment is placed at a different height $\beta h$, the corresponding frame $\xi\text{-}0\text{-}\eta_\beta$ is depicted in green and its size is $\Delta L_\beta$. {\bf A$'$:} The block undergoes a deformation from its natural shape (light blue): the size of the reference fiber changes to $\Delta l_\alpha$ or $\Delta l_\beta$. The energy cost associated to this deformation is the same whether the reference segment is placed at $\alpha h$ or $\beta h$. }
\label{figS1}
\end{figure}

 Consider the Fig.~\ref{figS1}. The undeformed elementary block is represented on the left A, and its shape upon deformation is displayed on the right A$'$. Let us assign to the block an integral planar reference system, where the $\xi$ and $\eta_\alpha$ axes define respectively the block's longitudinal and transverse directions. The origin of such reference system is placed at an height $\alpha h$ ($0\leq\alpha\leq1$) from the block's bottom surface, and on the left lateral block boundary.
 The quantity $\Delta L(\eta_\alpha)$ corresponds to the  length of a generic material line placed at the height $\eta_\alpha$, with $-\alpha h\leq \eta_\alpha\leq (1-\alpha)h$. By definition, the value of $\Delta L(\eta_\alpha=0)$ is the representative material line length $\Delta L_\alpha$. Now, a translation of the reference system  along the axis $\xi=0$ is equivalent to a linear change of variables:
 
 \be
 \eta_\beta=\eta_\alpha+(\alpha-\beta)h
 \label{ch_RF}
 \ee
 
 \noi where $\eta_\beta$ is the new axis pointing along the block transverse direction. However, the material line lengths $\Delta L(\eta_\alpha)$ have to be invariant under the transformation~(\ref{ch_RF}):

 \be
 \Delta L\left(\eta_\beta\right)=\Delta L\left(\eta_\alpha=\eta_\beta-(\alpha-\beta)h\right).
 \label{ch_length_undef}
 \ee

 \noindent It is also clear that in the reference system $\xi\text{-}0\text{-}\eta_\beta$, the representative material line has length $ \Delta L_\beta= \Delta L\left(\eta_\beta=0\right)= \Delta L\left(\eta_\alpha=(\beta-\alpha)h\right)$ and $-\beta h\leq \eta_\beta\leq (1-\beta)h$.
 
 Let us turn to the deformed configuration A$'$.
 The deformed longitudinal length $\Delta l$  follows the same law \eqref{ch_length_undef} under the shift of the reference system, i.e. 
 
 \be
 \Delta l\left(\eta_\beta\right)=\Delta l\left(\eta_\alpha=\eta_\beta-(\alpha-\beta)h\right).
 \label{ch_length_def}
 \ee

\noindent Therefore, since the extension is defined as $u=\Delta l-\Delta L$, thanks to Eqs.\eqref{ch_length_undef} and \eqref{ch_length_def} we have that the following equality holds

\be
 u\left(\eta_\beta\right)=u\left(\eta_\alpha=\eta_\beta-(\alpha-\beta)h\right).
 \label{ch_ext}
 \ee

\noi The strain energy of the block calculated in the   $\xi\text{-}0\text{-}\eta_\alpha$ reference frame is 

\be
\Delta E_\alpha=\frac{bY}{2}\int_{-\alpha h}^{(1-\alpha)h}\frac{u(\eta_\alpha)^2}{\Delta L(\eta_\alpha)} \ d\eta_\alpha .
    \label{E_block_RF}
\ee
    
\noi By applying the change of variables~(\ref{ch_RF}), the equality of the integral expression \eqref{E_block_RF}  $\Delta E_\alpha=\Delta E_\beta$ follows from the relations~(\ref{ch_length_undef}) and~(\ref{ch_ext}).

\section{Derivation of the strain energy from a flat undeformed configuration \label{sec:meth_derivation_flat}}   

\begin{figure}[h]
\centering
\includegraphics[width=6.1cm]{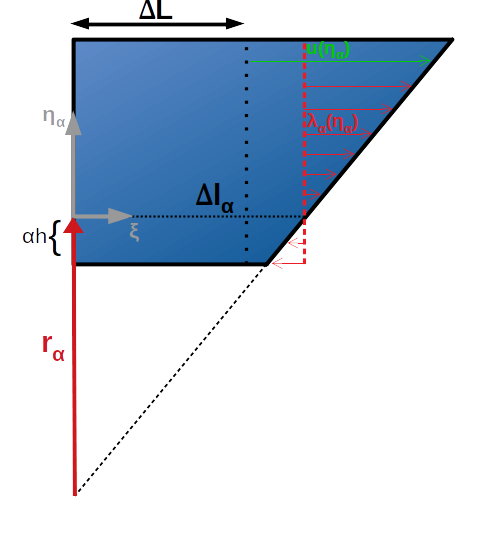}
\caption{Strain energy from a flat undeformed configuration. The block in its undeformed configuration has a longitudinal size of $\Delta L$ (dotted black vertical line). When deformed, the size of the representative fiber varies from $\Delta L$ to $\Delta l_\alpha$. In the reference frame integral with the block, any fiber undergoes a displacement $u(\eta_\alpha)$ (green arrows). At the same time, the size of any deformed  material fiber can be expressed as $ \Delta l(\eta_\alpha)= \Delta l_\alpha + \lambda_\alpha(\eta_\alpha)$, where $\lambda_\alpha(\eta_\alpha)$ are represented by red arrows.}
\label{figS2}
\end{figure}

 Let us consider the deformation of the elementary block presented in Fig.\ref{figS2}. The undeformed flat condition is depicted by a dotted black line, and it has the peculiarity that the longitudinal length is equal to $\Delta L$ for any choice of the representative segment. The plane integral reference frame is identified  by the $\xi$ and $\eta_\alpha$ axes, pointing respectively towards the block's longitudinal and transverse directions. The origin  is placed at an height $\alpha h$ ($0\leq\alpha\leq1$) from the block's bottom surface, and on the left lateral block boundary.
 Any fiber placed at an height $\eta_\alpha$ attains a length $\Delta l(\eta_\alpha)$ upon deformation, with $\Delta l(\eta_\alpha=0)=\Delta l_\alpha$. According to the geometrical construction  in Fig.\ref{figS2}, the Eq.\eqref{meth_elong_1_flat},
 is equivalent to

 \be
 \Delta l(\eta_\alpha)= \Delta L + u(\eta_\alpha)
 \label{meth_elong_2_flat}.
 \ee

\noi From Eqs.\eqref{meth_elong_2_flat}, \eqref{meth_elong_1_flat} and \eqref{meth_geom_rel} the elongation of any fiber can be expressed as

  \be
 u(\eta_\alpha)= \Delta l_\alpha - \Delta L + \eta_\alpha\frac{\Delta l_\alpha}{r_\alpha}
 \label{meth_elong_flat_final}.
 \ee

\noindent In this condition, the strain energy of the block  takes the form

\be
\Delta E_\alpha=\frac{bY}{2}\int_{-\alpha h}^{(1-\alpha)h}
\left[\frac{\Delta l_\alpha - \Delta L}{\Delta L} + \frac{\eta_\alpha}{r_\alpha}\frac{\Delta l_\alpha}{\Delta L}\right]^2 \Delta L\ d\eta_\alpha , 
    \label{meth_E_block_flat}
\ee

\noi where we have inserted the relation \eqref{meth_elong_flat_final}. Solving the integral and defining the strain as $\varepsilon_\alpha=\frac{\Delta l_\alpha - \Delta L}{\Delta L} $, we arrive at the expression

\be
\Delta E_\alpha = \frac{bY}{2}\left[h\,\Delta L\, \varepsilon_\alpha^2+h^2(1-2\alpha)\,\varepsilon_\alpha\frac{\Delta l_\alpha}{r_\alpha}+h^3\left(\frac{1}{3}-\alpha+\alpha^2\right)\frac{1}{\Delta L}\left(\frac{\Delta l_\alpha}{r_\alpha}\right)^2\right] \ . 
    \label{meth_E_block_flat}
\ee

\noindent It is clear that the value of $\alpha$ which guarantees the axial-bending uncoupling is the line of the centroids $\alpha_U=1/2$.

\begin{figure}[h]
\centering
\includegraphics[width=10.1cm]{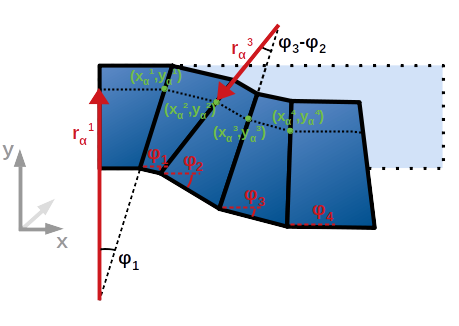}
\caption{Beam discrete strain energy from a flat undeformed configuration.  The polygonal, connecting the vertices of each deformed representative segments, is represented by a dotted black thick line.
In the laboratory frame, the vertices have planar coordinates $\left(x_\alpha^{(i)},y_\alpha^{(i)}\right)$ shown as green dots. The morphological change of each block is determined by the radius of curvature that can be positive (as $r_\alpha^1$) or negative (as $r_\alpha^3$). The connection between local longitudinal deformation of the representative fiber, $\Delta l_\alpha^{(i)}$, and  its radius of curvature $r_\alpha^{(i)}$, is encapsulated in the relation \eqref{meth_bending_angle}, where the bending angles $\varphi^{(i)}$ define the deflection of the deformed blocks  from the external $x$ axis.}
\label{figS3}
\end{figure}

The strain energy of the whole elastica is given by the sum over the block contributions, formally by $E_\alpha=\sum_{i=1}^N \Delta E_\alpha^{(i)}$. We aim at furnishing, however, its analytical expression in the laboratory frame (Fig.~\ref{figS3}). 

\noindent The single block's representative segment size $\Delta l_\alpha^{(i)}$ is a positive quantity, being $\Delta l_\alpha^{(i)}=\left|\mathbf{l}_\alpha^{(i)}-\mathbf{l}_\alpha^{(i-1)}\right|$. $\mathbf{l}_\alpha^{(i)}\equiv\left(x_\alpha^{(i)},y_\alpha^{(i)}\right)$ are  the vertices of the reference polygonal curve $l_\alpha$ in the lab reference system (Fig.~\ref{figS3}). The polygonal curve is defined as the ordered sequence of the representative segments $\Delta l_\alpha^{(i)}$. The quantity $r_\alpha^{(i)}$, on the other side,  can be positive or negative. This is required by the consistency of the Eq.(\ref{meth_geom_rel}). As a matter of fact, $\left|r_\alpha^{(i)}\right|$ becomes the radius of the osculating circle which locally approximates the reference segment in the continuum limit, and the  sign of $r_\alpha^{(i)}$ is assigned in the following way. It is clear that the intersection between the sidelines containing the block's sections lies on the axis $\xi=0$ of the local reference system integral with any block (see Fig.~\ref{figS6}). Hence, in this reference system the coordinates of the circle's center are defined as $\left(0,-r_\alpha^{(i)}\right)$: this establishes uniquely the sign of $r_\alpha^{(i)}$. Now, moving to the laboratory frame we find that the relation \eqref{meth_bending_angle} is always satisfied, with $\Delta \varphi^{(i)}=\varphi^{(i)}-\varphi^{(i-1)}$, and $\varphi^{(i)}$ corresponds to the bending angle between the $i-$th block and  the $x$ axis. In the small deflection limit, i.e. $\tan \Delta \varphi^{(i)}\simeq \Delta \varphi^{(i)}$, the discrete strain energy for the entire slender beam is therefore framed as

\be
E_\alpha=\frac{bY}{2}\sum_{i=1}^N \left[h\Delta L\, {\varepsilon^{(i)}_\alpha}^2-h^2(1-2\alpha)\,\varepsilon^{(i)}_\alpha\Delta\varphi^{(i)}+h^3\left(\frac{1}{3}-\alpha+\alpha^2\right)\,\frac{\Delta {\varphi^{(i)}}^2}{\Delta L}\right].
 \label{meth_E_dscrt-flat}
\ee

\noindent The bending measure is defined as

\be
\mu^{(i)}=\frac{-\Delta\varphi^{(i)}}{\Delta L}
\label{meth_bendging_meas_def1}
\ee

\noi or, thanks to \eqref{meth_bending_angle}, as

\be
\mu^{(i)}=\frac{\Delta l_\alpha^{(i)}}{\Delta L}\frac{1}{r_\alpha^{(i)}}
\label{meth_bendging_meas_def2}.
\ee

\noi Using these definitions, the energy~(\ref{meth_E_dscrt-flat}) takes the following form

\be
E_\alpha=\frac{bY\Delta L}{2}\sum_{i=1}^N \left[h\, {\varepsilon^{(i)}_\alpha}^2+h^2(1-2\alpha)\varepsilon^{(i)}_\alpha\mu^{(i)}+h^3\left(\frac{1}{3}-\alpha+\alpha^2\right)\mu^{(i)2}\right] 
 \label{meth_E_dscrt-flat}
\ee

Let us introduce two parametric expressions for the undeformed and deformed reference plane curves as $\mathscr{L}_\alpha:[0,L]\to \mathbbm{R}^2$ and  $\ell_\alpha:[0,L]\to \mathbbm{R}^2$ respectively. As a consequence, the Cartesian coordinates of the undeformed reference curve in the laboratory frame are

\be
\mathbf{L}_\alpha(s)=\left\{
\begin{array}{ccc}
X_\alpha(s) & = & s\\
Y_\alpha(s) & = & 0,
\label{meth_undef_coord-flat}
\end{array} 
\right.
\ee

\noi and those of the deformed curve $\ell_\alpha(s)$ are  $\mathbf{l}_\alpha(s)\equiv\left(x_\alpha(s),y_\alpha(s)\right)$. Taking an uniform partition of $[0,L]$, i.e. $0=s_0<s_1<s_2<\cdots<s_N=L$ such that $s_i-s_{i-1}=\Delta s \equiv\Delta L$ for any $i$, we obtain that the polygonal vertices are $\mathbf{l}_\alpha(s_i)=\mathbf{l}_\alpha^{(i)}$ and the local longitudinal strain $\varepsilon_\alpha^{(i)}$ is given by

\be
\varepsilon_\alpha(s_i)=\frac{\left|\Delta \mathbf{l}_\alpha(s_i)\right|}{\Delta s} -1
\label{meth_dscrt_strain-flat}.
\ee

\noindent Analogously, the bending angles at the polygonal vertices are $\varphi(s_i)=\varphi^{(i)}$. The continuum limit is taken by increasing $N$ until the length of the polygonal chain $l_\alpha$ approaches from below that of the curve $\ell_\alpha$, i.e. $\frac{\left|\Delta \mathbf{l}_\alpha(s_i)\right|}{\Delta s}\to \left|\ell_\alpha'(s)\right|$ as $\Delta s\to 0$. The curve derivative is defined as $\ell_\alpha'(s)\equiv\mathbf{t}_\alpha(s)$, where we have introduced the tangent of the curve $\mathbf{t}_\alpha(s)=\frac{d \mathbf{l}_\alpha(s)}{ds}$. Finally, if the continuum limit entails that $\frac{\Delta \varphi(s_i)}{\Delta s}\to \varphi'(s)$  and $\sum_{i=1}^N \Delta L\to \int_0^L ds$, by substitution of $\varepsilon_\alpha(s)=\left|\mathbf{t}_\alpha(s)\right|-1$ the energy (\ref{meth_E_dscrt-flat}) takes the form

\be
\mathscr{E}_\alpha = \frac{bY}{2}\int_{0}^{L}ds
\left\{h\, \left[\left|\mathbf{t_\alpha}(s)\right|-1\right]^2 -
h^2(1-2\alpha)\, \left[\left|\mathbf{t_\alpha}(s)\right|-1\right]\varphi'(s) +
h^3\left(\frac{1}{3}-\alpha+\alpha^2\right)\,\varphi'(s)^2\right\}.
\label{meth_E_ct_flat}
\ee

\section{Strain energy limiting cases: regaining the flat undeformed condition \label{sec:meth_E_limit-any-to-flat} }

In the present section we show how to recover the straight beam  strain energy~(\ref{meth_E_block_flat}), from the energy~(\ref{meth_E_block_any}) calculated from a generic
undeformed configuration. To this aim, it will be sufficient to study the behaviour of $F_\alpha$, $S_\alpha$ and $I_\alpha$ in the limit of $\frac{h}{\min[|R_0|,|R_1|]}\to0$. 

\noi Let us firstly express the relations~(\ref{meth_F_alpha})-(\ref{meth_I_alpha}) as

\be
F_\alpha=\left|R_\alpha\right|\ln\left(1+\frac{h}{\min[|R_0|,|R_1|]}\right)
\label{meth_F_alpha_compact},
\ee

\be
S_\alpha=R_\alpha\left[h-|R_\alpha|\ln\left(1+\frac{h}{\min[|R_0|,|R_1|]}\right)\right]
\label{meth_S_alpha_compact}
\ee

\noi and

\be
I_\alpha=R_\alpha\left[\left(\frac{1}{2}-\alpha\right)h^2-hR_\alpha+R_\alpha^2\ln\left(1+\frac{h}{\min[|R_0|,|R_1|]}\right)\right],
\label{meth_I_alpha_compact}
\ee

\noi Then we consider the condition $\frac{h}{\min[|R_0|,|R_1|]}\ll 1$ and expand the logarithm to the third order:

\be
\ln\left(1+\frac{h}{\min[|R_0|,|R_1|]}\right)\simeq \frac{h}{\min[|R_0|,|R_1|]}-\frac{h^2}{2\min[|R_0|,|R_1|]^2}+\frac{h^3}{3\min[|R_0|,|R_1|]^3}.
\label{meth_log_exp}
\ee

\noi Hence we get

\be
F_\alpha=\left\{
\begin{array}{ccc}
h+\frac{h^2}{R_0}\left(\alpha-\frac{1}{2}\right)+\frac{h^3}{R_0^2}\left(\frac{1}{3}-\frac{\alpha}{2}\right) &   & \min[|R_0|,|R_1|]=|R_0| \\
\ & & \\
h+\frac{h^2}{R_1}\left(\alpha-\frac{1}{2}\right)+\frac{h^3}{R_1^2}\left[\frac{1}{3}-\frac{(\alpha-1)}{2}\right] &   & \min[|R_0|,|R_1|]=|R_1| 
\end{array},
\right.
\label{meth_F_alpha_limit}
\ee

\be
S_\alpha=\left\{
\begin{array}{ccc}
h^2\left(\alpha-\frac{1}{2}\right)-\frac{h^3}{R_0}\left(\frac{1}{3}-\alpha+\alpha^2\right) &   & \min[|R_0|,|R_1|]=|R_0| \\
\ & & \\
h^2\left(\alpha-\frac{1}{2}\right)-\frac{h^3}{R_1}\left(\frac{1}{3}-\alpha+\alpha^2\right)&   & \min[|R_0|,|R_1|]=|R_1|, 
\end{array}
\right.
\label{meth_S_alpha_limit}
\ee

\be
I_\alpha=h^3\left(\frac{1}{3}-\alpha+\alpha^2\right) .
\label{meth_I_alpha_limit}
\ee

\noi By substitution of the former relations into (\ref{meth_E_block_any}), the Eq.~(\ref{meth_E_block_flat}) is  correctly reestablished.

\section{Macroscopic constitutive equations under change of material curve\label{sec:meth_Const_Eq_inv}}

When the natural state is flat, the strain energy function is defined as $W_\alpha=\frac{\Delta E_\alpha}{\Delta L}$, where $\Delta E_\alpha$ is given in  Eq.~(\ref{meth_E_block_flat}):

\be
W_\alpha=\frac{bY}{2}\left[h\, \varepsilon_\alpha^2+h^2(1-2\alpha)\varepsilon_\alpha\mu+h^3\left(\frac{1}{3}-\alpha+\alpha^2\right)\mu^2\right].
 \label{E_function-flat}
\ee

\noi The usual choice of the middle fiber as the representative medium ($\alpha=1/2$) yields the  expression commonly used in several contexts \cite{antman1968general,magnusson2001behaviour,oshri2016properties,atanackovic1998buckling}. However, for a generic choice of the representative fiber, the constitutive equations for the axial force and the bending moment are readily obtained:

\be
\begin{cases}
      N_\alpha=\frac{\partial W_\alpha}{\partial \varepsilon_\alpha}=bY\left[h\varepsilon_\alpha+h^2\mu\left(\frac{1}{2}-\alpha\right)\right]\\
      \ \\
      M_\alpha=\frac{\partial W_\alpha}{\partial \mu}=bY\left[h^2\varepsilon_\alpha\left(\frac{1}{2}-\alpha\right)+h^3\mu\left(\frac{1}{3}-\alpha+\alpha^2\right)\right].
    \end{cases}
    \label{meth_const_eq-flat}
\ee

\noi the axial force exerted on a material line $\alpha$ in (\ref{meth_const_eq-flat}) can be transformed into $N_\beta$  by applying the change of material line

\be
\varepsilon_\alpha=\varepsilon _\beta+h(\alpha-\beta)\mu.
\label{meth_ch_strain-dscrt-flat}
\ee

\noi It results immediately $N_\alpha=N_\beta$. 
The bending moment can be recast as

\be
M_\alpha=bY\left[h^2\varepsilon_\alpha\left(\frac{1}{2}-\alpha\right)+h^3\frac{\mu}{12}+h^3\mu\left(\frac{1}{2}-\alpha\right)^2\right].
\label{meth_bend_mom_flat}
\ee

\noindent Therefore, from the expression of the axial force we have

\be
M_\alpha=M_{1/2}-h\left(\alpha-\frac{1}{2}\right)N_\alpha.
\label{meth_bend_mom_flat_1}
\ee

\noindent Hence by subtracting the expression for $M_\beta$ from~(\ref{meth_bend_mom_flat_1}) and recalling the axial force invariance we have

\be
M_\alpha=M_{\beta}-h\left(\alpha-\beta\right)N_\beta.
\label{meth_bend_mom_flat_3}
\ee

The case of a general undeformed condition can be determined as follows. First we  notice how the strain transforms under change of material line

\be
\varepsilon_\alpha=\frac{\Delta L_\beta}{\Delta L_\alpha}\left[\varepsilon _\beta+h(\alpha-\beta)\mu_\beta\right].
\label{meth_ch_strain_1-any}
\ee

\noi Moreover, the reduced area, the reduced axial-bending coupling moments and the reduced moment of inertia change under material line transformation as

\be
F_\alpha=\frac{1}{\Delta L_\beta}\left[\Delta L_\beta-(\alpha-\beta)h\Delta \Phi\right]F_\beta
\label{meth_F_alpha_transf_semidscrt}
\ee

\be
S_\alpha=\frac{1}{\Delta L_\beta}\left[\Delta L_\beta-(\alpha-\beta)h\Delta \Phi\right]\left[S_\beta-(\alpha-\beta)hF_\beta\right]
\label{meth_S_alpha_transf_semidscrt}
\ee

\be
I_\alpha=\frac{1}{\Delta L_\beta}\left[\Delta L_\beta-(\alpha-\beta)h\Delta\Phi\right]\left[I_\beta-2(\alpha-\beta)hS_\beta+(\alpha-\beta)^2h^2F_\beta\right]
\label{meth_I_alpha_transf_semidscrt}.
\ee

\noi Inserting the previous relations  into

\be
\begin{cases}
      N_\alpha=\frac{\partial W_\alpha}{\partial \varepsilon_\alpha}=Y\left(F_\alpha\varepsilon_\alpha+S_\alpha\mu_\alpha\right)\\
      \ \\
      M_\alpha=\frac{\partial W_\alpha}{\partial \mu_\alpha}=Y\left(S_\alpha\varepsilon_\alpha+I_\alpha\mu_\alpha\right),
    \end{cases}
    \label{meth_const_eq-any}
\ee

\noi it easily turns out that the equality $N_\alpha=N_\beta$ holds also in this case. Moreover, plugging the same transformations
 into the second of Eqs.\eqref{meth_const_eq-any}, we recover the  Eq.~(\ref{meth_bend_mom_flat_3}) also in the case of generic initial conditions.

\section{The neutral fiber: from flat to ring and viceversa \label{sec:meth_circle_to_flat}}

\begin{figure}[h]
\centering
\includegraphics[width=9.1cm]{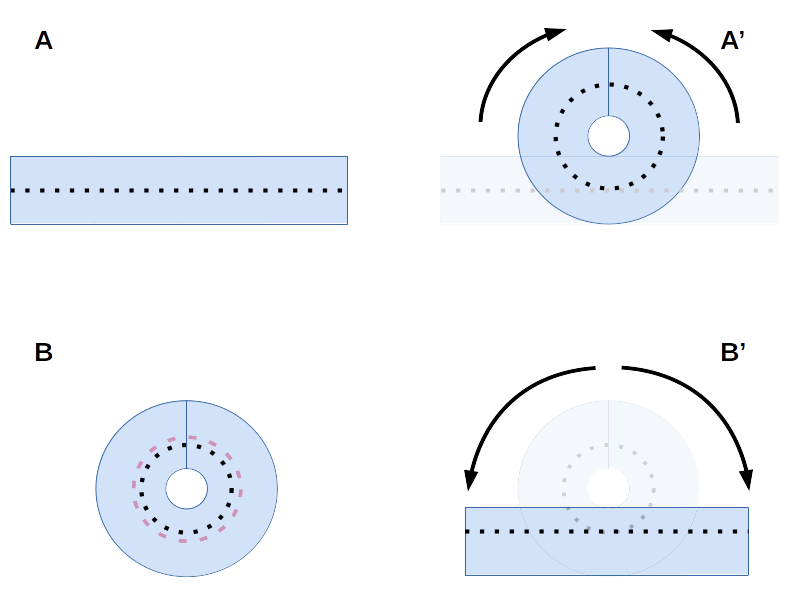}
\caption{}
\label{figS9}
\end{figure}

Let us consider the deformation depicted  in Fig.\ref{figS9}A-A$'$, where a slender beam of length $L$ is deformed into a circle. Let us take as the representative fiber the curve placed at an height $\alpha h$ from the bottom surface, so that the equation representative of the elastica undeformed configuration is

\be
\mathbf{L}_\alpha(s)=
\begin{cases}
X_\alpha(s)=s\\
\ \\
Y_\alpha(s)=\alpha h,
\end{cases}
\label{meth_eq_bar_undef}
\ee             

\noindent with $s\in[0,L]$. The tangent is expressed as 

\be
\mathbf{T}_\alpha(s)=
\begin{cases}
\frac{dX_\alpha(s)}{ds}=1\\
\ \\
\frac{dY_\alpha(s)}{ds}=0,
\end{cases}
\label{meth_eq_bar_tg_undef}
\ee   

\noindent and $\varPhi(s)=0$. By deforming the representative fiber into a circle of radius $r_\alpha$, we easily obtain 

\be
\mathbf{l}_\alpha(s)=
\begin{cases}
x_\alpha(s)=r_\alpha\cos\left(\frac{2\pi s}{L}\right)\\
\ \\
y_\alpha(s)=r_\alpha\sin\left(\frac{2\pi s}{L}\right),
\end{cases}
\label{meth_eq_circle_def}
\ee

\be
\mathbf{t}_\alpha(s)=
\begin{cases}
\frac{dx_\alpha(s)}{ds}=-\frac{2\pi r_\alpha}{L}\sin\left(\frac{2\pi s}{L}\right)\\
\ \\
\frac{dy_\alpha(s)}{ds}=\frac{2\pi r_\alpha}{L}\cos\left(\frac{2\pi s}{L}\right),
\end{cases}
\label{meth_eq_circle_tg_def}
\ee

\noindent and $\varphi(s)=\frac{\pi}{2}-\frac{2\pi s}{L}$. The energy necessary for the complete bending of the beam into the circle is given by \eqref{meth_E_ct_flat}

\be
\mathscr{E}_\alpha(L;r_\alpha) = \frac{bY}{2L}
\left[h\, \left(2\pi r_\alpha-L\right)^2 +
2\pi h^2(1-2\alpha)\, \left(2\pi r_\alpha-L\right) +
4\pi^2h^3\left(\frac{1}{3}-\alpha+\alpha^2\right)\right]
\label{meth_E_ct_flat_to_circle}.
\ee

\noindent Without loss of generality,  let us adopt the line of centroid as the representative material line, namely $\alpha=1/2$. We know that this choice has the only advantage of yielding the axial-bending uncoupling in Eq.\eqref{meth_E_ct_flat_to_circle}:

\be
\mathscr{E}_{1/2}(L;r_{1/2}) = \frac{bY}{2L}
\left[h\, \left(2\pi r_{1/2}-L\right)^2 +
\frac{\pi^2h^3}{3}\right]
\label{meth_E_ct_flat_to_circle_uncoupled}.
\ee

\noindent Nonetheless, if the transformation is such that $ r_{1/2}=\frac{L}{2\pi}$, i.e. the middle fiber maintains its length constant (zero strain condition), the energy has a minimum.   In other words, among all the possible deformations that transform a bar into a circle, that one which leaves unvaried the middle fiber (the neutral fiber) costs the minimum amount of work:

\be
\mathscr{E}_{1/2}\left(L;r_{1/2}=\frac{L}{2\pi}\right) = \frac{bY\pi^2h^3}{6L}
\label{meth_E_min_flat_to_circle}.
\ee

\noindent This minimum principle can be seen as the straightforward application of  Parent's principle $N_{1/2}=0$.

Now let us consider the opposite  situation, where a naturally curved beam is flattened into a bar as in Fig.\ref{figS9}B-B$'$. The undeformed configuration is given by 

\be
\mathbf{L}_\alpha(\theta)=
\begin{cases}
X_\alpha(\theta)=R_\alpha \cos(\theta)\\
\ \\
Y_\alpha(\theta)=R_\alpha \sin(\theta),
\end{cases}
\label{meth_eq_circle_undef}
\ee             

\noindent with $\theta\in[0,2\pi)$ being the internal parameter which is now adimensional, rather than having the dimension of an internal length. 

\be
\mathbf{T}_\alpha(\theta)=
\begin{cases}
\frac{dX_\alpha(\theta)}{d\theta}=-R_\alpha \sin(\theta)\\
\ \\
\frac{dY_\alpha(\theta)}{d\theta}=R_\alpha \cos(\theta),
\end{cases}
\label{meth_eq_circle_tg_undef}
\ee   

\noindent so that $\varPhi(\theta)=\frac{\pi}{2}-\theta$. On the other side the equation for the deformed bar of length $l$ is 
\be
\mathbf{l}_\alpha(\theta)=
\begin{cases}
x_\alpha(\theta)=\frac{l\theta}{2\pi}\\
\ \\
y_\alpha(\theta)=\alpha h,
\end{cases}
\label{meth_eq_bar_def}
\ee

\be
\mathbf{t}_\alpha(\theta)=
\begin{cases}
\frac{dx_\alpha(\theta)}{d\theta}=\frac{l}{2\pi}\\
\ \\
\frac{dy_\alpha(\theta)}{ds}=0,
\end{cases}
\label{meth_eq_circle_tg_def_1}
\ee

\noindent and $\varphi(\theta)=0$. Hence, 
the energy cost connected to such a transformation is


\be
\begin{split}
\mathscr{E}_\alpha(R_\alpha;l) = \pi bY 
\left\{\left(\frac{l}{2\pi}-R_\alpha\right)^2\ln\left(1+\frac{h}{R_0}\right)-2\left[h-R_\alpha\ln\left(1+\frac{h}{R_0}\right)\right]\left(\frac{l}{2\pi}-R_\alpha\right)+ \right. \\
\left. + \left[\left(\frac{1}{2}-\alpha\right)h^2-hR_\alpha+R_\alpha^2\ln\left(1+\frac{h}{R_0}\right)\right] \right\}.
\end{split}
\label{meth_E_ct_circle_to_flat}
\ee

\noindent In analogy to  the previous case, we choose the value of $\alpha$ which entails the axial-bending uncoupling, namely, according to Eq.\eqref{meth_alpha_NF_ct_1}, 

\be
\alpha_U=\frac{1}{\ln\left(1+\frac{h}{R_0}\right)}-\frac{R_0}{h}.
\label{meth_circle_to_flat_alpha_uncoupling}
\ee

\noindent Thanks to the fact that $R_\alpha=R_0+\alpha h$, from \eqref{meth_circle_to_flat_alpha_uncoupling} it results $R_{\alpha_U}=\frac{h}{\ln\left(1+\frac{h}{R_0}\right)}$. Hence the  Eq.\eqref{meth_E_ct_circle_to_flat} becomes

\be
\mathscr{E}_{\alpha_U}(R_0;l) = \pi bY 
\left\{\left(\frac{l}{2\pi}-\frac{h}{\ln\left(1+\frac{h}{R_0}\right)}\right)^2\ln\left(1+\frac{h}{R_0}\right)+ \left[\frac{1}{2}-\frac{1}{\ln\left(1+\frac{h}{R_0}\right)}\right]h^2 \right. +hR_0right\} 
\label{meth_E_ct_circle_to_flat_uncoupled}
\ee

\noindent Thus, it is possible to see that the minimum of energy necessary to flatten the ring is achieved only if the chosen uncoupling representative fiber keeps its length constant , i.e. $l=\frac{2\pi h}{\ln\left(1+\frac{h}{R_0}\right)}=2\pi R_{\alpha_U}$. Such amount of energy turns out to be

\be
\mathscr{E}_{\alpha_U}\left(R_0;l=\frac{2\pi h}{\ln\left(1+\frac{h}{R_0}\right)}\right)=\pi bY
\left\{R_0h+\left[\frac{1}{2}-\frac{1}{\ln\left(1+\frac{h}{R_0}\right)}\right]h^2\right\}
\label{meth_E_min_circle_to_flat}.
\ee

\noindent Again, the minimum of the energy is consistently required by the validity of  Parent's principle.

So far we have considered the generic situation where circle in panel B of Fig.\ref{figS9} and that in panel A$'$ are different. If we set the same dimensions for both of them, we have $R_0=\frac{L}{2\pi}-\frac{h}{2}$ that, inserted into the energy expression \eqref{meth_E_ct_circle_to_flat_uncoupled} yields

\be
\mathscr{E}_{\alpha_U}(L;l) = \pi bY 
\left\{\frac{l^2}{4\pi^4}\ln\left(\frac{L+\pi h}{L-\pi h}\right) + \frac{h}{\pi}\left(\frac{L}{2}-l\right)\right\}.
\label{meth_E_ct_circle_to_flat_uncoupled_2}
\ee

\noindent The amount of work needed to stretch the ring out, keeping constant the length of the neutral fiber, is 

\be
\mathscr{E}_{\alpha_U}\left(L;l=\frac{2\pi h}{\ln\left(\frac{L+\pi h}{L-\pi h}\right)}\right)=\pi bYh
\left\{\frac{L}{2\pi}-\frac{h}{\ln\left(\frac{L+\pi h}{L-\pi h}\right)}\right\}
\label{meth_E_min_circle_to_flat_2}.
\ee

\noindent Conversely if we want to stretch the ring keeping constant the line of the centroids, therefore it is sufficient to replace $l=L$ into the expression \eqref{meth_E_ct_circle_to_flat_uncoupled_2}:

\be
\mathscr{E}_{\alpha_U}(L;l=L) = \pi bY 
\left\{\frac{L^2}{4\pi^4}\ln\left(\frac{L+\pi h}{L-\pi h}\right) - \frac{hL}{2\pi}\right\}.
\label{meth_E_ct_circle_to_flat_uncoupled_3}
\ee